\documentclass[a4paper,10pt]{article}
\usepackage[style=numeric]{biblatex}
\usepackage{braket}
\usepackage[english]{babel}
\usepackage[T1]{fontenc}
\usepackage[utf8]{inputenc}
\usepackage{graphicx}
\usepackage{mleftright}
\usepackage{amsmath}
\usepackage{fullpage}
\usepackage{setspace}
\usepackage{mathrsfs}
\usepackage{amsfonts}
\usepackage{subcaption}
\usepackage{chngcntr}
\usepackage[colorlinks=true,linkcolor=blue,urlcolor=blue,citecolor=blue]{hyperref}
\usepackage{physics}
\usepackage{float}
\usepackage{svg}
\usepackage{copyrightbox}
\usepackage{url}
\usepackage{mathtools}
\DeclarePairedDelimiter\floor{\lfloor}{\rfloor}
\usepackage{outlines}
\usepackage{comment}

\begin{document}


\begin{titlepage}
\doublespacing
\centering \includegraphics[scale=1]{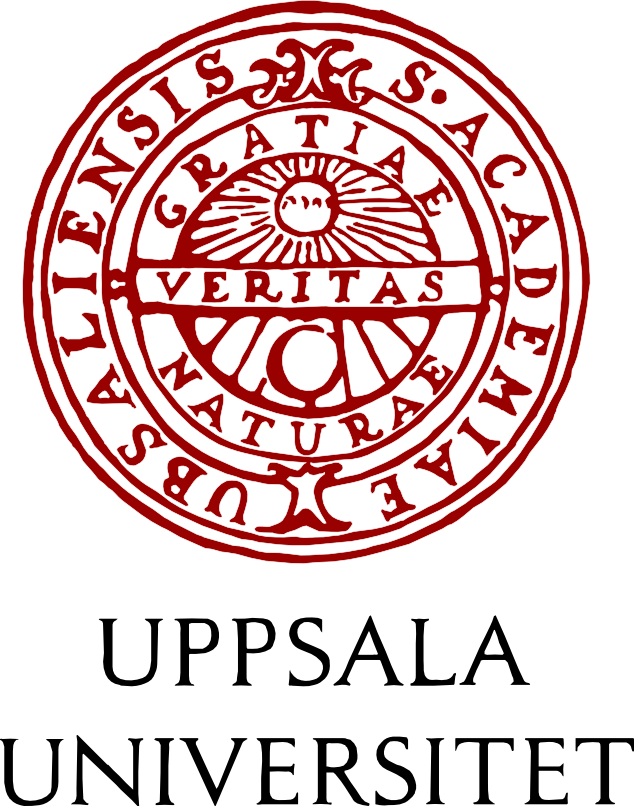}\\	
\vspace{3 cm}
{\LARGE {\bf Quantum Computing: Implementing Hitting Time for Coined Quantum Walks on Regular Graphs}}  
\vspace{2 cm} \\

Ellinor Wanzambi, Stina Andersson

\vspace{3 cm}
Uppsala\\ \today    
\end{titlepage}
\section*{Acknowledgments}
During the writing of this Master Thesis, we have received much support from different people. This thesis would not have been possible without their help, and we would like to offer our gratitude towards them. Firstly, we cannot begin to express our thanks to Researcher Dr. Dhinakaran Vinayagamurthy and Research Engineer Dhiraj Madan at IBM Research, who has helped us continuously throughout the entire process. We are highly grateful for the time you have spent to help and support us. Both of you have played decisive roles in deciding the topic for the thesis, and it is thanks to your knowledge and valuable advice that we have been able to complete it. We would also want to extend our sincere thanks to Mikael Haglund, CTO at IBM Sweden, for your guidance and support. Your profound belief in our abilities has encouraged us greatly during this process. Lastly, we want to give many thanks to Associate Professor Carl Nettelblad for reviewing our thesis and giving us many practical suggestions. You have always been willing to give of your time to help us, which has made the process of writing this thesis considerably easier.

\newpage

\begin{abstract}

In recent years, quantum walks have been widely researched and have shown exciting properties. One such is a quadratic speed-up in hitting time compared to its classical counterpart. In this paper, we design a quantum circuit for the MNRS algorithm, which finds a marked node in a graph with a quantum
walk, and use it to find a hitting time for the marked nodes in the walk. We do this by implementing the circuit on IBM quantum simulators and show that the execution on a noise-free simulator results in hitting times that agree with the theoretical expectations. We also run the algorithm on a mock backend that simulates the noise of the IBM Melbourne computer. As expected, the noise has an extensive impact on the output, resulting in outcomes far from the noise-free simulation.

\end{abstract}
\newpage
\tableofcontents
\newpage



\section{Introduction}

In October 2019, Google published a report in \textit{Nature}, announcing that they had achieved quantum supremacy \cite{q_supremacy}. In only 200 seconds, the quantum computer managed to do calculations that would take the best classical computer 10 000 years to complete. This was a major milestone in quantum computing, and there is much more to come. IBM has revealed that their target for the end of 2023 is a quantum device with more than 1000 qubits \cite{ibm_blog}. To put this in context: already with 300 qubits, a quantum computer can do more calculations \textit{simultaneously} than there are particles in the universe \cite{chalmers}. \\

Alongside hardware development, progress has also been made in the area of quantum algorithms. Two of the most famous algorithms are Shore's algorithm for integer factorization \cite{shor}, and Grover's algorithm for searching an unsorted database \cite{grover}. Both have shown a significant speed-up compared to their classical counterparts. Something that has turned out to be a key in many quantum algorithms is \textit{quantum walks}, which is the quantum equivalence of a classical Markov chain, i.e., a walk without memory. Quantum walks are useful in areas such as searching \cite{mnrs, search}, node ranking in networks \cite{node_ranking2, node_ranking} and element distinctness \cite{element_distinctness}. The two most popular models for discrete-time quantum walks are coined quantum walks and Szegedy's quantum walk. Coined quantum walks are walks on the vertices of a graph. Szegedy's walk, on the other hand, is a walk on the edges of a bipartite double cover of the original graph. These two models are equivalent under certain circumstances, namely when the coin used in the coined walk is a Grover coin \cite{wong_article}. Most research on quantum walks has been on their theoretical properties, but there has been a few papers on circuit design and implementation. A 2016 paper by Loke and Wang \cite{loke_wang} shows circuits for implementing Szegedy's walk for a few different graphs and a paper by Douglas and Wang \cite{douglas_wang} presents some circuits for coined quantum walks. \\

The MNRS algorithm \cite{mnrs} is a quantum search algorithm that finds a marked element in a graph with a quantum walk. In this paper, we design a quantum circuit for the MNRS algorithm, and use it to find the walk's \textit{hitting time}. The hitting time for a (quantum or classical) random walk is the minimum number of steps required to reach a marked vertex in a graph. We also give quantum circuits for walks with Grover coin on a hypercube, 2-dimensional lattice, complete bipartite graph, and complete graph, which are the graphs that we use in the quantum walk search algorithm. Finally, we implement the search algorithm with all four graphs on IBM quantum simulators and compare the resulting hitting times to what has been derived from theory. This paper is structured as follows. Section \ref{sec:theory} provides an introduction to quantum computing, quantum walks and the quantum walk search algorithm. In section \ref{sec:method}, we present the quantum circuits for the graphs and the quantum walk search algorithm. The results are given and discussed in sections \ref{sec:result} and \ref{discussion} respectively. Section \ref{recommendation} gives some recommendations for future work and the conclusions are found in section \ref{conclusion}.

\section{Theory} \label{sec:theory}
\subsection{Hilbert space}
The state of a quantum system can be described by a vector in its associated \textit{Hilbert space}, which we also refer to as the \textit{state space}. A Hilbert space is a complex vector space with an inner product \cite{qwalks}.

\subsection{Quantum computing}
Computers make use of bits and Boolean logic gates to store and manipulate information. In a classical computer, bits are binary and can be either 0 or 1. Quantum computers have logic gates that are based on the principles of quantum mechanics. Instead of acting on regular binary bits, these gates operate on so-called \textit{quantum bits}, or \textit{qubits}. Qubits have the property that they cannot only be in one of the two binary states; they can also be in combinations of both states simultaneously, called a \textit{superposition}. As a consequence, this allows us to develop new, faster algorithms that outperform their classical counterparts. In quantum computing, we start with a set of qubits, all in state $\ket{0}$, and apply a sequence of gates to manipulate the qubits, before we do a \textit{measurement} to obtain the result \cite{q_firstbook}. 

\subsubsection{Qubits}
In the same way that bits are crucial in classical computers, qubits are fundamental in quantum computing. A classical bit is always in a single state, $\ket{0}$ or $\ket{1}$, while a qubit can also be in a so-called mixed state:

\begin{equation}
    \ket{\psi} = \alpha \ket{0} + \beta \ket{1},
\label{eq:qubit_state}
\end{equation}

i.e. a linear combination of the \textit{computational basis states} $\ket{0}$ and $\ket{1}$. This is called \textit{superposition}. The coefficients $\alpha$ and $\beta$ are complex numbers and are called the \textit{amplitudes} of state. It is impossible to directly observe the state of a qubit in superposition. Instead, we need to apply a \textit{measure gate}. When we measure a qubit, we will not get a mixed state as in equation \eqref{eq:qubit_state}; the resulting state will be either $\ket{0}$ or $\ket{1}$ and we will not know the state of the qubit before the measurement. At the time of measurement, we say that the qubit \textit{collapses} to state $\ket{0}$ or $\ket{1}$, and the probabilities for each state are given by the square of the coefficients in \eqref{eq:qubit_state}. That is, the qubit collapses to $\ket{0}$ with probability $|\alpha|^2$ and to $\ket{1}$ with probability $|\beta|^2$ \cite{qcomputation}. From this follows that the coefficients in \eqref{eq:qubit_state} must satisfy 
\begin{equation}
    |\alpha|^2 + |\beta|^2 = 1.
\end{equation}

The states $\ket{0}$ and $\ket{1}$ can also be expressed by unit vectors in a 2-dimensional Hilbert space, $\mathcal{H}^2$, 

\begin{equation}
    \ket{0} = \begin{bmatrix} 1 \\ 0 \end{bmatrix} , \;
    \ket{1} = \begin{bmatrix} 0 \\ 1 \end{bmatrix}. 
\label{eq:vec_two_qubits}
\end{equation}

If we have two qubits, the computational basis is the four states $\ket{00}$, $\ket{01}$, $\ket{10}$, $\ket{11}$. In general, the computational basis consist of the possible distinct states a qubit, or a set of qubits, can collapse to in the measurement. Similarly as in the one qubit case, two qubits can exist in superposition of the computational basis,

\begin{equation}
    \ket{\psi} = \alpha_{00} \ket{00} +  \alpha_{01} \ket{01} +  \alpha_{10} \ket{10} +  \alpha_{11} \ket{11},
\end{equation}

where $|\alpha_{ij}|^2$ denotes the probability that the qubit collapse in state $\ket{ij}$ \cite{qcomputation}. Also here the squares of the amplitudes must sum to 1. We can generalize this idea to $n$ qubits; an $n$ qubit system has $2^n$ computational basis states, which can be represented by unit vectors in the $2^n$-dimensional Hilbert space. For the state to be a valid quantum state the amplitudes must satisfy $\sum\limits_{i \in x} |a_i|^2 = 1$, where $x$ is the set of all $2^n$ computational basis states.  \\

The state of a quantum system of $n$ qubits can be expressed as follows: let $\ket{\psi_1}$,..., $\ket{\psi_n}$ be the states of each of the $n$ qubits. Then, the state of the quantum system is the tensor product of the individual states, i.e., $\ket{\psi_1} \otimes \cdot \cdot \cdot \otimes \ket{\psi_n}$ \cite{qwalks}. \\

Now consider the state 

\begin{equation}
    \ket{\psi} = \frac{\ket{00} + \ket{11}}{\sqrt{2}},  
\end{equation}

which collapses to $\ket{00}$ or $\ket{11}$ with equal probability $1/2$. Even though the qubits are in superposition, we will know the state of the second qubit by only measure the first since the only possible states are that the two qubits collapse to the same state. That is, if we measure the first qubit and get $\ket{1}$ we know, without applying a measure gate to the second qubit, that it also will collapse to $\ket{1}$. Similarly, we can obtain the state of the first qubit by measuring only the second. We say that the two qubits are \textit{entangled}, or that they are in \textit{entangled states} \cite{Qiskit-Textbook}. On the contrary, the state 

\begin{equation}
    \ket{\psi} = \frac{\ket{00} + \ket{01}}{\sqrt{2}}.
\end{equation}

is not entangled since we do not know the state of the second qubit by only measure the first; it can be both $\ket{0}$ and $\ket{1}$. Entangled qubits cannot be written as the tensor product of the individual states. Thus, in a quantum state of $n$ qubits, if $m$ qubits can be represented by $\ket{\psi_1} \otimes \cdot \cdot \cdot \otimes \ket{\psi_m}$, $m \leq n$, the $m$ qubits are \textit{not} entangled \cite{qnotes}.

\subsubsection{Quantum gates}
Operations on quantum states are called \textit{gates}. Let $\ket{\psi}$ be a quantum state. We can transform it into state $\ket{\psi'}$ by applying the gate $U$ as follows, 

\begin{equation}
    \ket{\psi'} = U\ket{\psi}.
\end{equation}

A gate is represented by a unitary matrix, which means that the norm, and thereby the probabilities, of the state vector is unaffected by the operation. This is required for the quantum state after the transformation to be valid since the probabilities must sum to 1 both before and after the operation. It is also necessary that all gate matrices are unitary since one requirement on the gates is that they are reversible; combining a gate immediately followed by its conjugate transpose should return the original quantum state before the gates were used. A quantum gate operating on $n$ qubits has degree $2^n$. \\

One of the most simple gates is the \textit{NOT} or \textit{(Pauli-) X} gate. It inverts a qubit, so $\ket{0}$ becomes $\ket{1}$ and $\ket{1}$ converts to $\ket{0}$. The matrix representation of the NOT gate is

\begin{equation}
    U_{\text{NOT}} =  \begin{bmatrix}
                        0 & 1 \\
                        1 & 0
                    \end{bmatrix}.
\end{equation}

Using the vector notation of the computational basis states for two qubits introduced in \eqref{eq:vec_two_qubits} we can see that the NOT gate inverts the qubits as expected \cite{qgates}:

\begin{align}
    U_{\text{NOT}} \ket{0} &= \begin{bmatrix}
                        0 & 1 \\
                        1 & 0
                    \end{bmatrix} \begin{bmatrix} 1 \\ 0 \end{bmatrix}  = \begin{bmatrix} 0 \\ 1 \end{bmatrix} = \ket{1} \\
    U_{\text{NOT}} \ket{1} &= \begin{bmatrix}
                        0 & 1 \\
                        1 & 0
                    \end{bmatrix} \begin{bmatrix} 0 \\ 1 \end{bmatrix}  = \begin{bmatrix} 1 \\ 0 \end{bmatrix} = \ket{0}.
\end{align}

To put a qubit into superposition we can apply the \textit{Hadamard gate}, H for short, which puts the qubit into an equal superposition of the computational basis. It puts $\ket{0}$ into the state $\frac{\ket{0} + \ket{1}}{\sqrt{2}}$ and $\ket{1}$ in $\frac{\ket{0} - \ket{1}}{\sqrt{2}}$. The matrix is given by 

\begin{equation}
    U_H = \frac{1}{\sqrt{2}} \begin{bmatrix}
                        1 & 1 \\
                        1 & -1
                    \end{bmatrix}.
\label{eq:hadamard_gate}
\end{equation}

The $-1$ in the matrix makes the gate reversible.
Both the NOT gate and the Hadamard gate act on one qubit but there are gates that operate on multiple qubits. One such gate is the \textit{controlled-NOT}, or \textit{CNOT}, gate. The CNOT gate acts on two qubits - the \textit{control} qubit and the \textit{target} qubit. If the control qubit is $\ket{1}$, it flips the state of the target. In other words, it applies the NOT gate to the target qubit if the control is in state $\ket{1}$. This can be written in terms of the XOR operator: 

\begin{equation}
    \ket{\psi_1, \psi_2} \mapsto \ket{\psi_1, \psi_2 \oplus \psi_1}, 
\end{equation}

and in matrix notation \cite{qgates}:

\begin{equation}
    U_{\text{CNOT}} = \begin{bmatrix}
    1 & 0 & 0 & 0 \\
    0 & 1 & 0 & 0 \\
    0 & 0 & 0 & 1 \\
    0 & 0 & 1 & 0
    \end{bmatrix}.
\end{equation}

We can extend this idea and use more control qubits. A \textit{Toffoli gate} has two control qubits and one target. Similarly to the CNOT gate, we can express it with the XOR operator:

\begin{equation}
    \ket{\psi_1, \psi_2, \psi_3} \mapsto \ket{\psi_1, \psi_2, \psi_3 \oplus \psi_1  \psi_2}.
\end{equation}

It is also possible to implement an $n$-qubit Toffoli gate with the same idea as above. We then have $n-1$ control qubits and one target. \\

We can implement multiple gates in sequence; this is called a \textit{quantum circuit}. Assume that we have a unitary operator $U$ that consists of two other unitary operators $U_1$ and $U_2$, $U=U_1 U_2$. If we apply $U$ to some state $\ket{\psi}$ we get the following expression: 

\begin{equation}
    U\ket{\psi} = U_1 U_2 \ket{\psi}.
\end{equation}

\begin{figure}[h!]
\centering
\includegraphics[width=3cm]{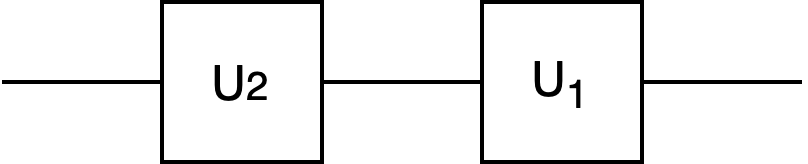}
\caption{Implementation of $U=U_1 U_2$.}
\label{fig:rev}
\end{figure}

As we see in the equation, we first apply $U_2$ on $\ket{\psi}$ followed by $U_1$. When we implement this circuit, we must place the gates in the order of application. That is, $U_2$ before $U_1$. This applies to all quantum circuits. Figure \ref{fig:rev} shows the correct implementation for $U$. 


\subsubsection{Phases}
One thing that distinguishes quantum computers from their classical counterparts is \textit{phases}. In quantum computing, we work with both \textit{global phase} and \textit{relative phase}. Let us consider the states $e^{i \theta} \ket{\psi}$ and $\ket{\psi}$. They differ by $e^{i \theta}$, which is the \textit{global phase factor}, and $\theta$ is defined as the phase. The global phase does not have any physical meaning and does not change the measurement probabilities. That is, in terms of measurement, they are equivalent. The relative phase of a system has a different meaning. Consider the matrix for the Hadamard gate given in equation \eqref{eq:hadamard_gate}. If we apply it to a $\ket{0}$ state, we get the mixed state 

\begin{equation}
    \frac{\ket{0} + \ket{1}}{\sqrt{2}},
\label{eq:h_0}
\end{equation}

and if we apply it to state $\ket{1}$ we obtain 

\begin{equation}
    \frac{\ket{0} - \ket{1}}{\sqrt{2}}.
\label{eq:h_1}
\end{equation}

We can see that the amplitude for state $\ket{0}$ is the same in both \eqref{eq:h_0} and \eqref{eq:h_1} but the amplitude for $\ket{1}$ differs; in \eqref{eq:h_0} it is $\frac{1}{\sqrt{2}}$ and in \eqref{eq:h_1} it is $-\frac{1}{\sqrt{2}}$. We say that the states differ in \textit{relative phase} with relative phase factor $-1$. If we have two arbitrary amplitudes $a_1$ and $a_2$ we say that they differ in relative phase if there exists a $\theta \in \mathbb{R}$ such that $a_1 = e^{i \theta} a_2$ \cite{qcomputation}. Now, let us apply another Hadamard gate on the states \eqref{eq:h_0} and \eqref{eq:h_1}. We start with the state in \eqref{eq:h_0} and obtain the following: 

\begin{align}
    H\Big(\frac{\ket{0} + \ket{1}}{\sqrt{2}}\Big) &= \frac{1}{\sqrt{2}} H(\ket{0}) + H(\ket{1}) \\
    &= \frac{1}{\sqrt{2}} \Big(\frac{\ket{0} + \ket{1}}{\sqrt{2}} + \frac{\ket{0} - \ket{1}}{\sqrt{2}} \Big) \\
    &= \frac{1}{2} \cdot 2\ket{0} \\
    &= \ket{0}.
\end{align}

Following the same logic when we apply the Hadamard gate on the state \eqref{eq:h_1} we get 

\begin{equation}
    H \Big(\frac{\ket{0} - \ket{1}}{\sqrt{2}} \Big) = \ket{1}.
\end{equation}

We can see that terms are canceling out because of difference in the relative phases. This property, called \textit{interference}, makes it possible to ``force'' the outcome of a qubit towards a wanted state, something that is widely used in many quantum algorithms \cite{q_firstbook}.

\subsection{Quantum oracle}
In quantum algorithms we sometimes want to give an input $x$ and receive an output $f(x)$ for some function $f$. In the algorithms, we usually assume that there exists some black box implementation of this function, referred to as an \textit{oracle}. We know that quantum gates must be reversible, i.e., given an output $f(x)$ we are able to uniquely identify the input $x$. This is not true in general; it only works when every input $x$ gives a unique output $f(x)$. However, we can make the transformation reversible by having a copy of the input in the output. One of the most common form of oracles is \textit{Boolean oracles}, which we define by the unitary evolution 

\begin{equation}
    U_f \ket{x, \Bar{0}} = \ket{x, f(x)}.
\label{eq:bool_oracle}
\end{equation}

As we see in equation \eqref{eq:bool_oracle}, the first register keeps track of the input state $\ket{x}$ and does not change throughout the transformation. This register consists of the number of qubits required to represent $\ket{x}$. The second register encodes the output. The input $\ket{\Bar{0}}$ denotes the state in which all qubits are $\ket{0}$ and $\ket{f(x)}$ is the binary representation of the output. The second register must have as many qubits as needed to represent $\ket{f(x)}$. \\

Another form of oracles is the \textit{phase oracle}, where $f(x)$ is a bit value, either 0 or 1. A phase oracle shifts the relative phase of $\ket{x}$ by $-1$ if $f(x)=1$. We define a phase oracle as \cite{Qiskit-Textbook}

\begin{equation}
    P_f \ket{x} = (-1)^{f(x)} \ket{x}.
\label{eq:phase_oracle}
\end{equation}

\subsection{Quantum Fourier transform}
The discrete Fourier transform takes a vector in $\mathcal{C}^N$ and maps it to a vector in $\mathcal{C}^N$. If we let the input vector consist of the elements $x_0, ..., x_{N-1}$ we can define the elements of the output, i.e. the transformed data, as 

\begin{equation}
    y_k \equiv \frac{1}{\sqrt{N}} \sum_{j=0}^{N-1} x_j e^{2 \pi i j k / N}, \; k=0,...,N-1.
\end{equation}

The \textit{quantum Fourier transform} is similar to the discrete one but acts on a quantum state $\ket{x} = \sum_{i=0}^{N-1} x_i \ket{i}$ and converts it into another quantum state $\ket{y} = \sum_{k=0}^{N-1} y_k \ket{k}$:

\begin{equation}
    \sum_{i=0}^{N-1} x_i \ket{i} \mapsto \sum_{k=0}^{N-1} y_k \ket{k},
\end{equation}

where $x_j$ are the amplitudes of the non-transformed data and $y_k$ are the amplitudes after the Fourier transform. We can define the quantum Fourier transform as a linear operator as follows, 

\begin{equation}
    \ket{j} \mapsto \frac{1}{\sqrt{N}} \sum_{j=0}^{N-1} e^{2 \pi i j k / N} \ket{k}, \; k=0,...,N-1.
\end{equation}

The quantum Fourier transform is a unitary transformation and, hence, it can be implemented in a quantum computer. Let $R_k$ denote the gate 

\begin{equation}
    R_k \equiv  \begin{bmatrix}
                    1 & 0\\
                    0 & e^{2 \pi i / 2^k}
                \end{bmatrix}.
\end{equation}

\begin{figure}[h!]
\centering
\includegraphics[width=15cm]{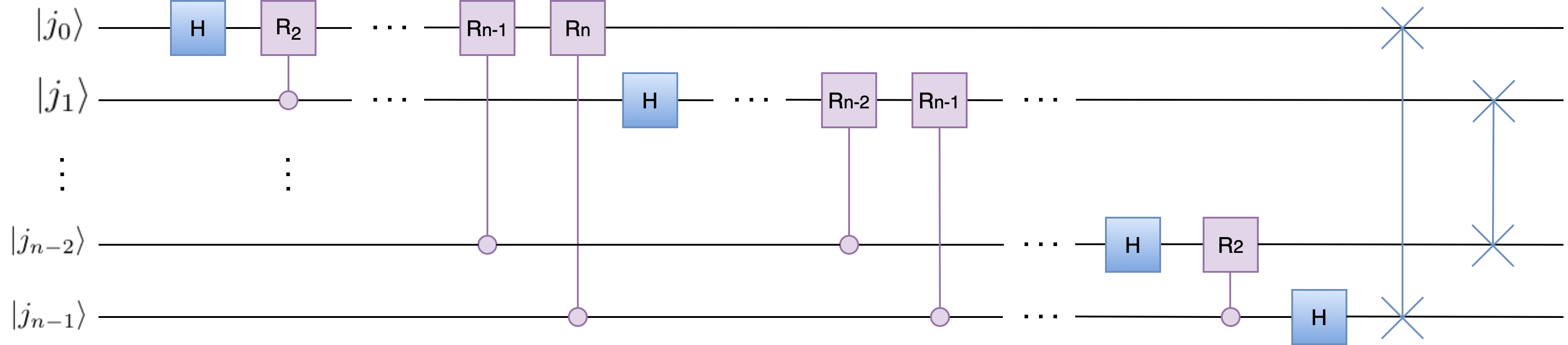}
\caption{Circuit implementation of the quantum Fourier transform. }
\captionsetup{font={footnotesize,it}}
\caption*{Source: \cite{qcomputation}}
\label{fig:qft}
\end{figure}

Then the quantum circuit for the quantum Fourier transform is given in Figure \ref{fig:qft} \cite{qcomputation}. \\

\subsubsection{Phase estimation}
\begin{figure}[h!]
\centering
\includegraphics[width=10cm]{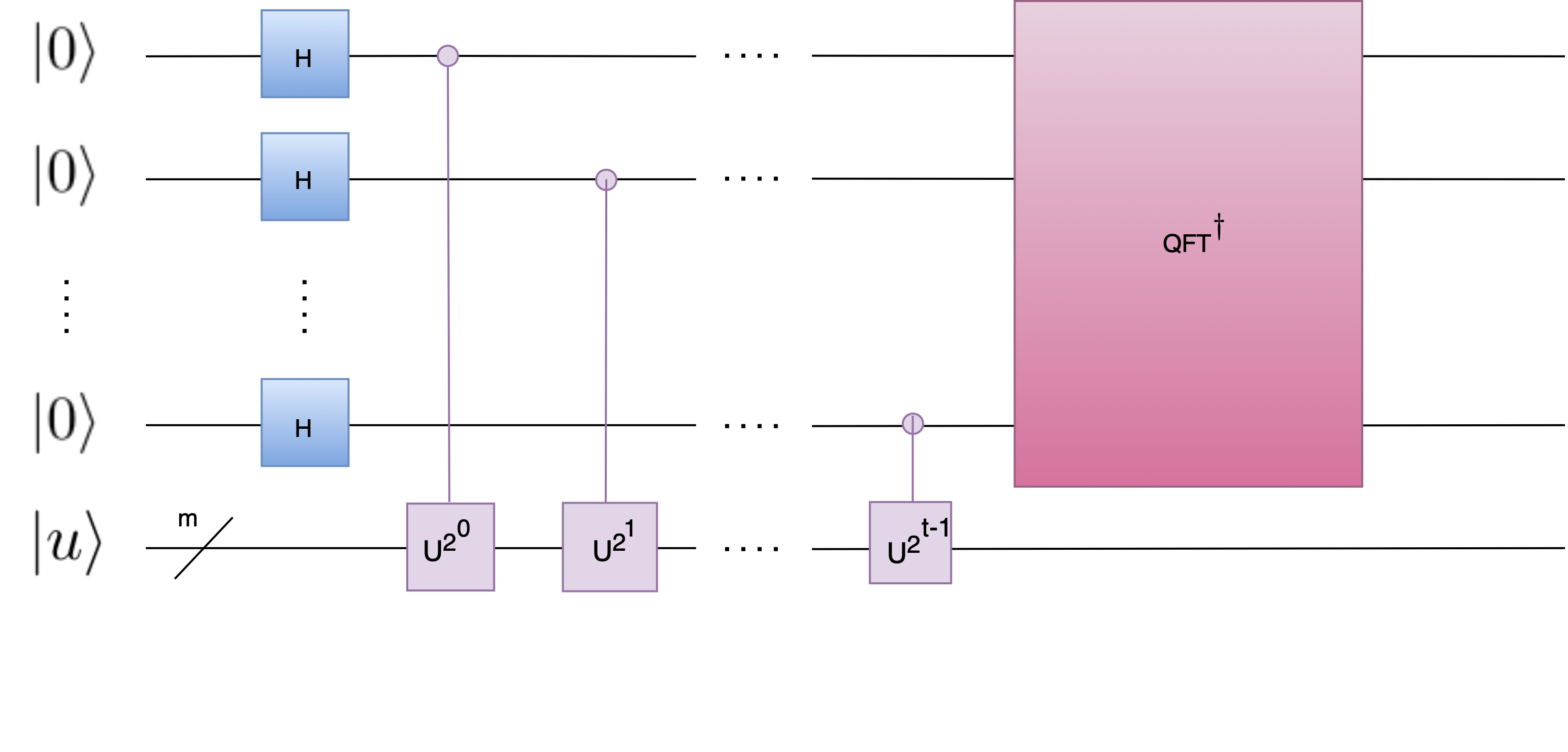}
\caption{Quantum circuit for the phase estimation. The '/' denotes $m$ qubits.}
\label{fig:phase_est}
\end{figure}

One important property of the quantum Fourier transform is that it enables \textit{phase estimation}, an approximation of the eigenvalues of a unitary operator. Phase estimation is a general procedure that turns out to be a key ingredient in many quantum algorithms. Assume that $U$ is a unitary operator that has an eigenvector $\ket{u}$ with corresponding eigenvalue $e^{2 \pi i \theta}$. Given that $\theta$ is unknown, the goal of the phase estimation is to find the value of $\theta$. To do this, we have two registers. The first one has $t$ qubits, all starting in state $\ket{0}$, and we choose $t$ depending on how many digits of accuracy we want to have and with what probability we want to succeed with the estimation. The other register starts in state $\ket{u}$, and has as many qubits as required to store $\ket{u}$. The phase estimation is done in two steps. We start by applying Hadamard gates on the first register, along with a sequence of controlled-$U$ gates raised to successive powers of two on the second register. After this, we apply the inverse quantum Fourier transform on the first register. We get the phase estimation by measuring the first register. To obtain the inverse transform, we reverse the circuit of the original quantum Fourier transform given in Figure \ref{fig:qft} \cite{qcomputation}. The full implementation of the phase estimation is presented in figure \ref{fig:phase_est}. 

\subsection{Grover's algorithm}
Consider the problem of searching for an element in an unstructured dataset. Given that the set contains $N$ elements, this can be done in $O(N)$ in a classical computer. With a quantum computer, we can solve the same problem using only $O(\sqrt{N})$ operations. The algorithm that enables this is called \textit{Grover's search algorithm} or just \textit{Grover's algorithm}. The algorithm makes use of an oracle function $f(x)$, such that $f(x)=1$ if and only if $x$ is a solution to the search problem. Otherwise, $f(x)=0$. Let $x_0$ be the element we are searching for. The oracle is represented by a unitary operator $U_f$ such that 

\begin{align}
    U_f = -\ket{x} \: &\text{if} \: x=x_0, \nonumber \\
    U_f = \ket{x} \: &\text{if} \: x \neq x_0.
\end{align}

\begin{figure}[h!]
\centering
\includegraphics[width=10cm]{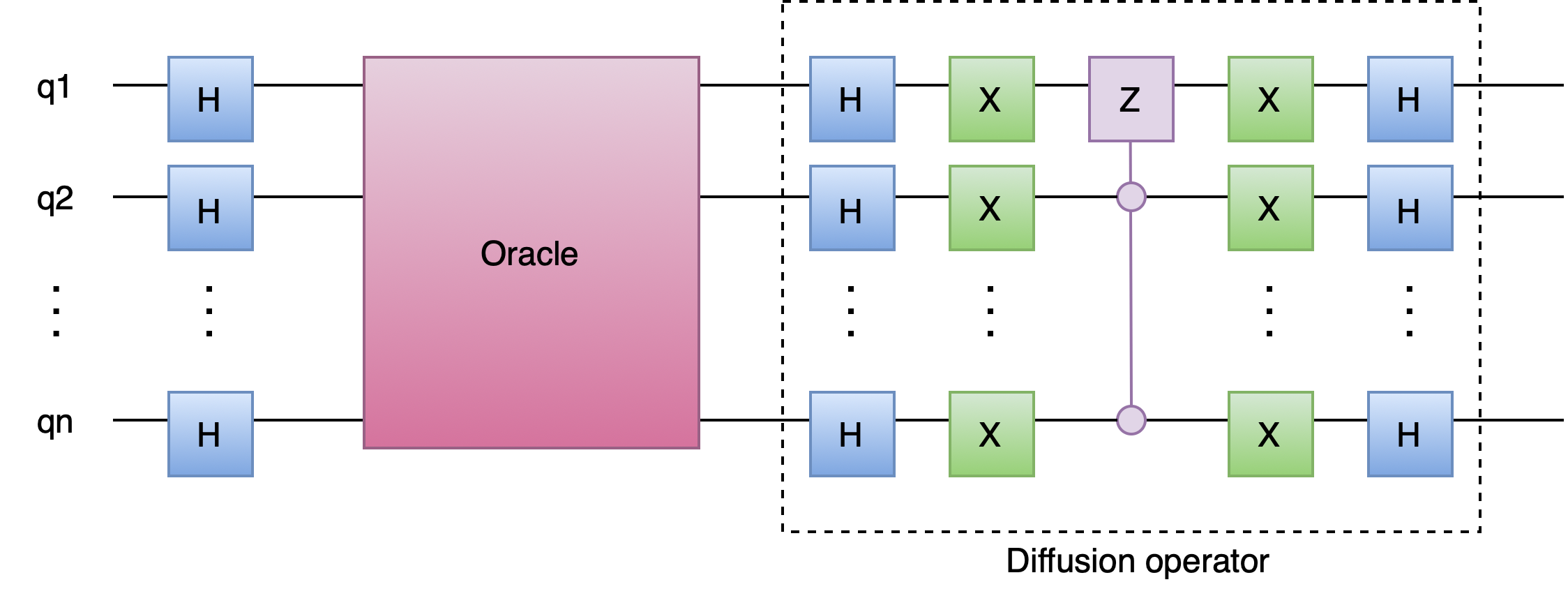}
\caption{Quantum circuit for one step of Grover's algorithm.}
\label{fig:grover}
\end{figure}

Looking at equation \eqref{eq:phase_oracle} we see that the unitary operator above represents a phase oracle, i.e., it marks the solutions by shifting the phase of $\ket{x}$. The algorithm begins with applying a uniform superposition to all qubits, $\ket{s} = \frac{1}{\sqrt{N}} \sum_{x=0}^{N-1} \ket{x}$. After this initial step the circuit has equal probability to collapse to each state in the computational basis. Now we define the \textit{diffusion operator}, $U_d$, as $2\ket{s}\bra{s} - I$. Then the evolution operator for one step of the Grover algorithm is $U = U_d U_f$. After the initial condition, we apply $U$. If we do so $\floor{\frac{\pi}{4}\sqrt{N}}$ times before measuring the result will be $x_0$ with probability $1-\frac{1}{N}$ \cite{qwalks}. The one-step circuit for Grover's algorithm is presented in Figure \ref{fig:grover}. 

\subsection{Qiskit} \label{sec:qiskit} 
Qiskit is an open-source framework for quantum computing provided by IBM. Qiskit makes it possible to write quantum programs using Python and running it on simulators or actual quantum prototypes. Qiskit can be used both on a local computer and in IBM Quantum Experience, a cloud-based quantum computing service \cite{Qiskit-Textbook}. \\

The quantum simulator provided by IBM allows running circuits with ideal quantum states without noise. This is something that is not possible on real quantum computers, which in general suffers from much noise. However, we can also simulate this noise in a quantum simulator, making it possible to achieve results closer to those of an actual quantum device. Simulators also provide other features that are not possible on today's hardware, such as up to 5000 qubits and the ability to run larger circuits \cite{q_simulator}. \\

As we have mentioned, there is no way to know the state of a quantum system. All we can do is a measurement that outputs the states the qubits collapsed to, but we will never know the state of the system before the measurement. However, in a quantum simulator, we can do just this by looking at the \textit{state vector}. The state vector is a vector that summarizes the state of the system at a defined time point. With a real quantum computer, we will not have this information since the qubits operate in a closed system that we cannot observe. Having access to the state vector means that we can inspect the system's state and thereby know the probabilities of collapsing to all possible states at any point in time. This is something that will be useful for us when we run the quantum walk search algorithm \cite{Qiskit-Textbook}.

\subsubsection{State vector vs measurement output}
Two plots that can easily be confused are the state vector plot and the measurement output plot. As mentioned above, the state vector summarizes the state of the quantum system at a certain time point and is something that we can observe when we run a circuit on a quantum simulator. A state vector plot is a histogram that shows the probabilities that the circuit collapses to each state in the computational basis at that time point. When we run a full quantum circuit (on either a simulator or a quantum prototype), the program executes multiple times. The number of execution times is called \textit{shots}, and one common value is $2^{10}$. In each execution, the full program runs, including the measurement. The measurement plot shows the frequency of the output states after all shots. So, the bars in a measurement plot show $\frac{\text{number of times circuit collapses to }x_i}{\text{total number of shots}}$ for each state $x_i$ in the computational basis. 

\subsubsection{IBM Melbourne quantum computer}
\begin{figure}[h!]
    \centering
    \includegraphics[width=12cm]{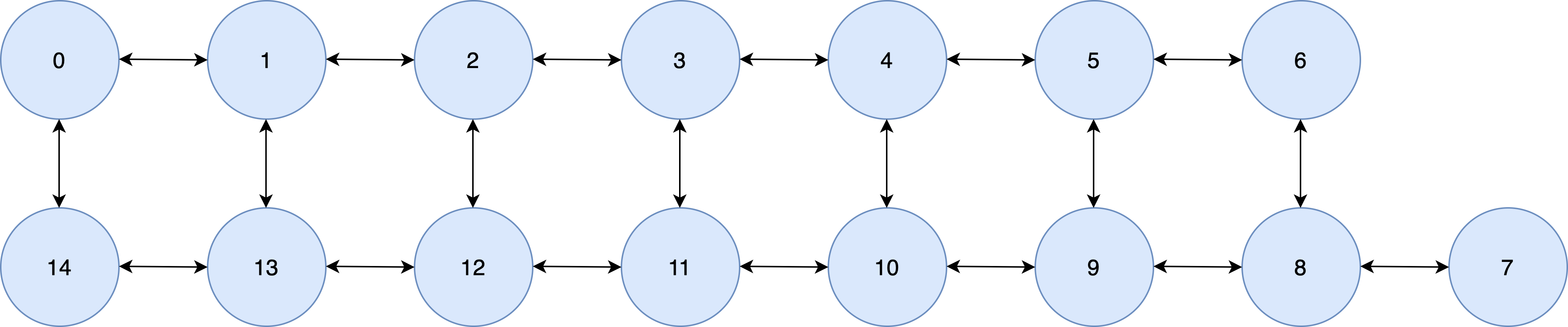}
    \caption{Qubit architecture in the Melbourne quantum computer. }
    \label{fig:melbourne_architecture}
\end{figure}

Melbourne is a quantum device provided by IBM. It is accessible to the public via IBM Quantum Experience and has 15 qubits, more than any of the other publicly accessible quantum computers on the platform. It has five \textit{basis gates}, CNOT, I, $R_Z$, SX, and NOT gate. This means that the circuit will be transpiled to an equivalent sequence of these gates in during compilation, since these are the only gates that are executable in the physical quantum device. Figure \ref{fig:melbourne_architecture} shows the architecture of the Melbourne computer \cite{melbourne}. How the qubits are physically connected in the quantum device affects the compilation of the circuit. Operations between qubits that are not physically connected in general require more gates than operations between physically linked qubits. 

\subsubsection{Mock backends}
As we have previously mentioned, it is possible to run quantum simulators with simulated noise. If we use so-called \textit{mock backends}, we can simulate the noise model of a \textit{specific} quantum device. By doing so, we get an approximation of the results that we would get if the program was executed on the actual quantum computer. These models build on a finite set of input parameters linked to the gates average error rates. Hence, we can only approximate the errors on the real device. These mock backends are also referred to as \textit{fake backends} \cite{mock_backends}. If we, for instance, write fake Melbourne, we refer to a mock backend with a noise model associated with the Melbourne computer.

\subsection{Quantum computers and noise}
\label{sec:noiseee}
\begin{figure}[h!]
\centering
\parbox{5cm}{
\includegraphics[width=6cm]{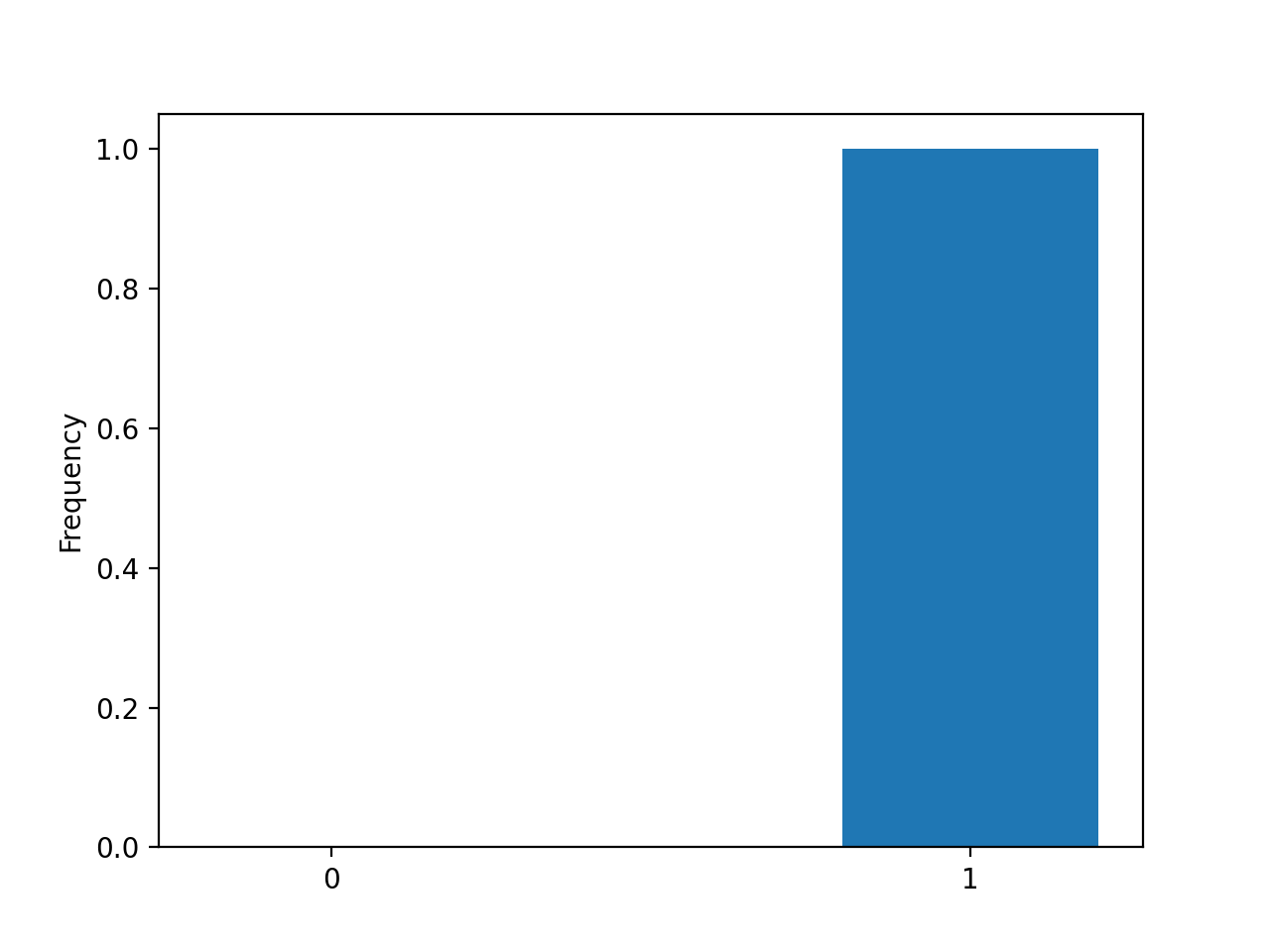}
\caption{Qiskit simulator output after applying a NOT gate to a qubit in state $\ket{0}$.}
\label{fig:freq_simulator}}
\qquad
\begin{minipage}{5cm}
\includegraphics[width=6cm]{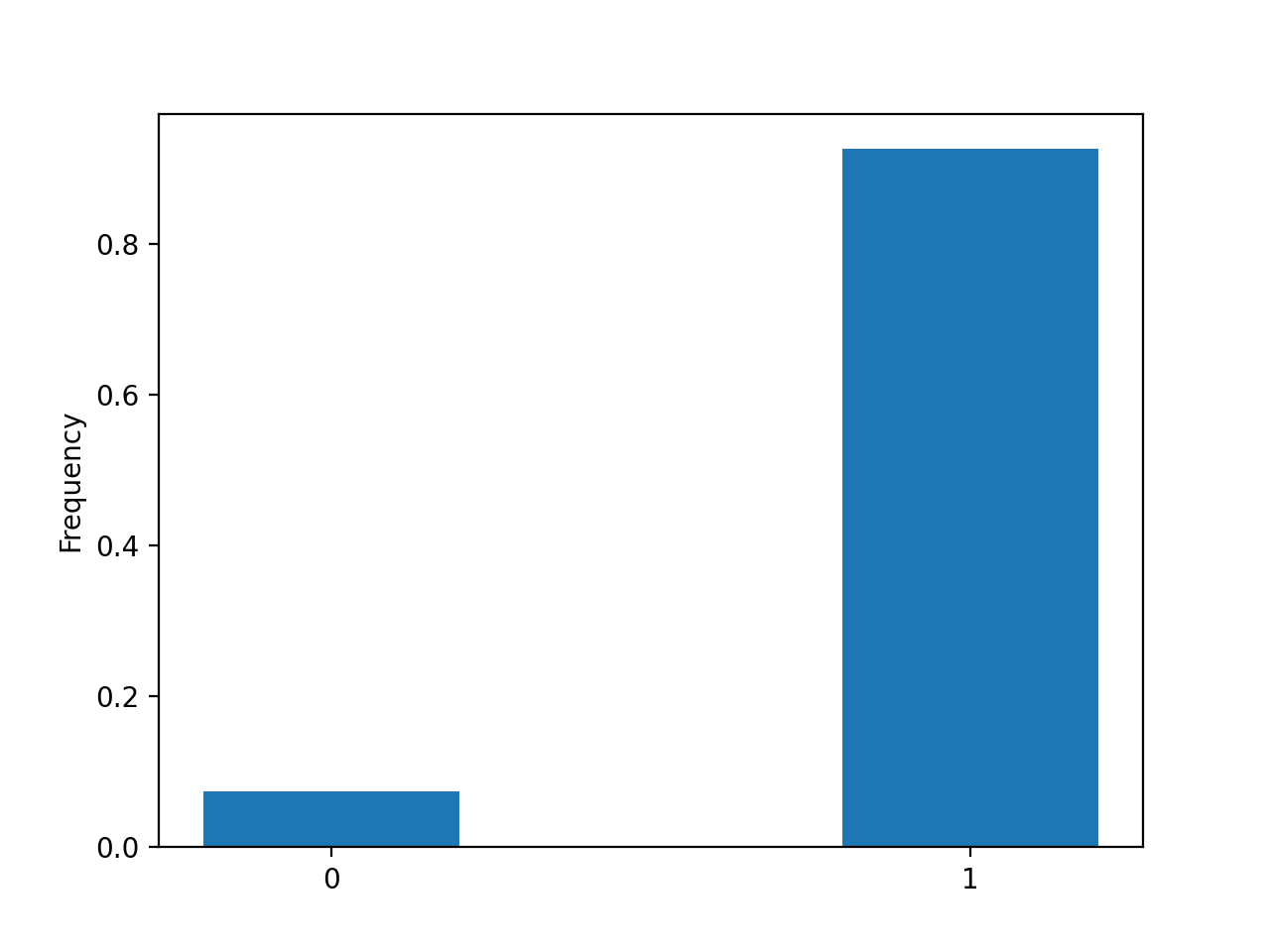}
\caption{Qiskit quantum prototype output after applying a NOT gate to a qubit in state $\ket{0}$.}
\label{fig:freq_computer}
\end{minipage}
\end{figure}

As mentioned in section \ref{sec:qiskit}, Qiskit offers the possibility to run circuits on quantum simulators as well as actual quantum computers. In Figures \ref{fig:freq_simulator} and \ref{fig:freq_computer}, we see the outputs after running the same simple circuit on a noise-free quantum simulator and a quantum computer. The circuit applies a NOT gate to a $\ket{0}$ qubit. As we see in Figure \ref{fig:freq_simulator}, the simulator always outputs $\ket{1}$, which is the expected output state. However, in Figure \ref{fig:freq_computer} the qubit collapses to state $\ket{0}$ roughly $7\%$ of the times. This due to \textit{noise} in the quantum computer. \\

Ideally, qubits operate in a closed environment without interaction with the surroundings, until we measure them. Today there is no way to fully accomplish this. Another problem is that qubits are not stable enough to always hold their quantum effects; they sometimes accidentally lose their quantum states, and we have not yet found a way to achieve complete control of this. Because of this, there is a noise associated with each qubit \cite{noisy_qubits}. Altogether, this means that the more qubits we have in our circuit, the larger the noise. 

\subsection{Hardware-aware circuits}
The qubits in a quantum computer are usually not fully connected, as is the case with the previously mentioned IBM Melbourne quantum computer. The compiler does not only transpile a circuit to consist of basis gates entirely; it also transpiles the circuit to fit the qubit architecture in the physical hardware. This is achieved with swap operators, which swaps qubits such that qubits that interact with each other are directly connected. The reason is that we can only apply a multi-qubit operation to qubits that are directly connected. So, if we apply a multi-qubit gate to qubits that are not connected in the physical hardware, the transpiler must swap the qubits to use the gate. \\

All these swap operators required for multi-qubit gates may lead to a transpiled circuit that is much deeper than the original circuit. Since the noise increases with the number of gates, relatively small circuits might be transpiled into much larger circuits that suffer heavily from noise. Since the noise from just one gate is quite large, the noise in a  sequence of gates will quickly become such an issue that the output from the circuit is worthless. Not only does the noise increase, the execution time for these transpiled circuits also increases \cite{Niu_2020}. \\

We can decrease the depth of a transpiled circuit by making it hardware-aware. One way to do this is by constructing the circuit such that we minimize the number of swaps in the transpiled circuit. We can do this by letting qubits that often interact with each other be qubits closely connected in the physical hardware architecture. For example, if the same two qubits have many two-qubit gates acting on them, we arrange these two qubits to be right next to each other. 

\subsection{Classical Markov chains} \label{sec:classical_markov_chain}
A Markov chain is a stochastic process often used to model real-life processes. It consists of a set of states and associated transition probabilities, which describe the probabilities to move between the states in each time step. This report considers discrete-time walks, implying that there are discrete, enumerable, time steps. Markov chains satisfy the \textit{Markov property}, meaning that the conditional probability distribution of future states only depends on the present state. If $X$ is a Markov chain, we can mathematically describe the Markov property as

\begin{equation}
    P(X_t = x_t | X_{t-1} = x_{t-1}, ... , X_0 = x_0) = P(X_t=x_t | X_{t-1} = x_{t-1}).
\end{equation}

\begin{figure}[h!]
\centering
\includegraphics[width=10cm]{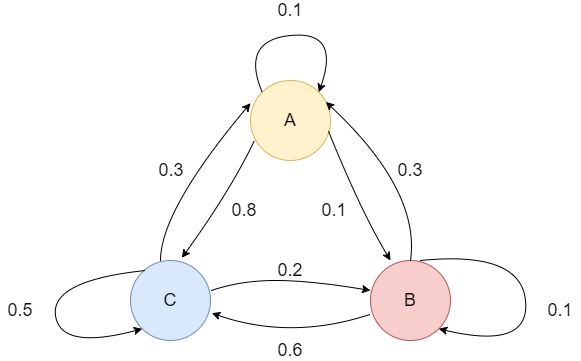}
\caption{A directed graph showing a Markov chain with three states represented by nodes. The probabilities of moving between the different states are shown as the directed edges between the states.}
\label{fig:markov_example}
\end{figure}

An example of a Markov chain is given in Figure \ref{fig:markov_example}. As we see in the graph, the probability to move to another state depends only on the current state. The sum of all edges going out from a state has to be equal to 1. \\

The \textit{transition matrix} for the Markov chain in Figure \ref{fig:markov_example} is
 
\begin{equation}
P=
\begin{pmatrix}
0.1 & 0.3 & 0.3\\
0.1 & 0.1 & 0.2 \\
0.8 & 0.6 & 0.5
\end{pmatrix}.
\label{eq:matrix_example}
\end{equation}

The transition matrix shows how likely an entity is to move between states in one time step. We can obtain the probabilities for multiple time steps by multiplying the transition matrix with itself, i.e., $P^t$. Let vector $v_0$ represent the initial state distribution. $v_0 = (1, 0, 0)^T$ means that it starts at state A with probability 1, $v_0 = (\frac{1}{3}, \frac{1}{3}, \frac{1}{3})^T$ means that it starts in state A, B, or C with equal probability. Given an initial probability vector $v_0$ and a transition matrix $P$, the state probabilities after $t$ time steps can be expressed as:

\begin{equation}
    v_t = P^tv_0.
\end{equation}

We say that a Markov chain is \textit{ergodic} if it, for all pair of states $i, j$, exists a $T$ such that there is a positive probability of being in state $j$ for all $t>T$ given that we start in state $i$. An ergodic Markov chain will, regardless of the initial state, converge to a stationary distribution. That is, for an ergodic Markov chain there exists a probability distribution $v_c$ such that

\begin{equation}
    v_c = Pv_c.
\end{equation}

If we select the initial distribution $v_0$ such that $v_0 = v_c$,  walking a time step on the graph would not change the probability distribution of the walker being at a certain state. \textit{Mixing time} is the time it takes to get close to this stationary distribution. Let the distance between two distributions $p$ and $q$ on the finite state space $\Omega$ be

\begin{equation}
    D(p, q) = \frac{1}{2} \sum_{x \in \Omega} |p_x - q_x|,
\end{equation}
where
\begin{equation}
   \sum_{x \in \Omega} p_x = \sum_{x \in \Omega} q_x = 1.
\end{equation}
Let the stationary distribution on $\Omega$ be $\pi$ and the transition matrix of the Markov chain on $\Omega$ be P. Also let

\begin{equation}
    d(t) = \underset{x \in \Omega}{max}   [D(P^t(x,\cdot), \pi)].
\end{equation}

Then the mixing time $t_{mix}$ for some threshold value $\epsilon$ can be defined as \cite{mixing_time} \cite{qwalks}:

\begin{equation}
    t_{mix}(\epsilon) = min\{t: d(t) \leq \epsilon \}.
\end{equation}

The \textit{hitting time} of a Markov chain is the expected number of time steps needed to reach a certain state given an initial state. Let $p_{xx'}$ be the likelihood that state $x'$ is reached for the first time after time $t$ after starting in state $x$. Then hitting time $H_{xx'}$ from state $x$ to state $x'$ can be found with \cite{qwalks}:
\begin{equation}
    H_{xx'} = \sum_{t=0}^{\inf} p_{xx'}(t). 
\end{equation}

\subsection{Quantum walks}
Quantum walks are the quantum computing equivalent of a Markov chain. As described previously, quantum computers have the property that they can be in multiple quantum states at the same time. For quantum walks, this means that, as long as one does not measure the qubits during the walk, the walker will take all possible paths simultaneously until it is measured. This leads to quantum walks having different properties compared to classical Markov chains. One such is that quantum walks do not converge to a stationary distribution \cite{qwalks}. 
\\

There are multiple ways to quantize a classical Markov chain. In this thesis, we will use two of the most common models, \textit{quantum walk with coins} and \textit{Szedgedy's quantum walk}. 

\subsubsection{Quantum walk with coins} \label{sec:q_walk_coins}
This model, also called \textit{the coin model}, consists of two quantum states and two operators. The states are the \textit{position state}, which represents the position of the walker, and the \textit{coin state}, which represents how the walker should move in the next time step. The position state can be represented as a vector in Hilbert space $\mathcal{H}_P$, while the coin state can be expressed as a vector in the Hilbert space $\mathcal{H}_C$. Given this, the quantum space representing the entire walker is $\mathcal{H} = \mathcal{H}_C \otimes \mathcal{H}_P$. One simple quantum walk is the walk on the infinite integer line. The computational basis for the position state is then $\{\ket{j} : j \in  \mathbb{Z} \}$, and the computational basis for the coin state is $\{\ket{0}, \ket{1}\}$, depending on whether the walker goes to the right or to the left on the integer line \cite{qwalks}.
\\

The two operators in the model are the \textit{coin operator}, $C$, and the shift operator, $S$. The coin operator acts on $\mathcal{H}_C$ during each time step and puts the walker in superposition. For the quantum walk on the integer line this means that it will walk both to the left and to the right at the same time. There are several different coin operators but a common one is the \textit{Hadamard coin}. The Hadamard coin is the most common choice for simple one-dimensional quantum walks
\begin{equation}
    H = \frac{1}{\sqrt{2}}
\begin{bmatrix}
1 & 1 \\
1 & -1 
\end{bmatrix}.
\end{equation}

Another common coin is the Grover coin. If the Hilbert space of the coin is defined as $\mathcal{H}^n$, then the Grover coin matrix is defined as \cite{qwalks}:

\begin{equation}
    G = 
    \begin{bmatrix}
\frac{2}{n} -1 & \frac{2}{n} & \ldots & \frac{2}{n}\\
\frac{2}{n} & \frac{2}{n} - 1  & \ldots & \frac{2}{n} \\
\vdots & \vdots & \ddots & \vdots \\
\frac{2}{n} & \frac{2}{n} & \ldots & \frac{2}{n} -1
\end{bmatrix}.
\end{equation}

The shift operator, $S$, will act on $\mathcal{H}_P$ and can also act on $\mathcal{H}_C$. The shift operator will move the walker to the next position, for the walk on a line the shift operator will act in the following way:
\begin{equation}
    S \ket{0}\ket{j} = \ket{0}\ket{j+1},
\end{equation}

\begin{equation}
    S \ket{1}\ket{j} = \ket{1}\ket{j-1}.
\end{equation}

With the shift operator and the Hadamard coin operator defined as above, we can represent one step of the quantum walk on the integer line as the unitary operator $U$ given by
\begin{equation}
    U = SH.
\end{equation}
The same evolution operator $U$ can be represented with any coin $C$ as
\begin{equation}
    U = SC.
\label{eq:coin_op}
\end{equation}

We can also express the quantum state $\ket{\psi}$ after $t$ time steps can be represented as
\begin{equation}
    \ket{\psi (t)} = U^t \ket{\psi(0)},
\end{equation}
where $\ket{\psi(0)}$ is the initial state \cite{qwalks}. 
\\

The quantum space for any finite regular graph where it is possible to color the edges of the graph using $d$ different colors can be generalized to $\mathcal{H} = \mathcal{H}^d \otimes \mathcal{H}^N $. $\mathcal{H}^N$ is the Hilbert space representing a graph with N states and $\mathcal{H}^d$ represents the coin space where d is the degree of each state node.
The computational basis will then be \cite{qwalks}
\begin{equation}
    \{\ket{a,v}, 0 \leq a \leq d-1, 0 \leq v \leq N-1 \}.
\end{equation}

The downside of using the quantum walk with coins is that it is most suitable for regular graphs and can be quite difficult to generalize to other types of graphs \cite{qnotes}. An alternative quantum walk model more convenient to use on other types of graphs is reviewed in the following section. 



\subsubsection{Szedgedy's quantum walk model} \label{sec:szegedy_model}
While a coined walk is a walk on the graph's nodes, Szegedy's walk is a walk on the edges on a double-cover of the original graph. As described in section \ref{sec:classical_markov_chain}, a classical discrete-time random walk is represented by a transition matrix $P$. For any $N$-vertex graph with $N \times N$ transition matrix $P$, we can define the corresponding discrete-time quantum walk as a unitary operation on the Hilbert space $\mathcal{H}^N \otimes \mathcal{H}^N$. Let $P_{jk}$ define the probability of making a transition from state $j$ to $k$. Before we define the walk, we define the normalized states 

\begin{equation}
    \ket{\psi_j} := \sum_{k=1}^N \sqrt{P_{kj}} \ket{j,k}, \; j=1,...,N
\end{equation}

and the projection onto ${\ket{\psi_j}}:j=1,...,N$

\begin{equation}
    \Pi := \sum_{j=1}^N \ket{\psi_j} \bra{\psi_j}.
\label{eq:sz_pi}
\end{equation}

We also introduce the shift operator S:

\begin{equation}
    S := \sum_{j,k=1}^N \ket{j,k} \bra{k,j}.
\label{eq:sz_s}
\end{equation}

With $S$ and $\Pi$ defined as above we can introduce one step of the discrete-time quantum walk:

\begin{equation}
    U := S(2 \Pi - 1),
\label{eq:sz_op}
\end{equation}

where $(2 \Pi - 1)$ is the reflection operator. We also define $t$ steps of the walk as $U^t$ \cite{qnotes}.

\subsubsection{Equivalence of coined and Szegedy's model}
It is known that a coined walk with Grover coin is equivalent to Szegedy's quantum walk. For more details we refer to this article \cite{wong_article} by Thomas G. Wong, where he also shows the equivalence between the operators in the two models.

\subsubsection{Hitting time}
A \textit{hitting time} for a (classical or quantum) random walk is the minimum number of steps required to reach one or a set of marked states in the state space. We define the quantum walk hitting time as the number of steps needed to maximize the probability to collapse to one of the marked states. For hitting time, quantum walks provide a quadratic speed-up compared to its classical counterpart \cite{hitting_time_chiang}. 


\subsection{Quantum walk search algorithm}
\label{section:qwa}
The quantum walk search algorithm solves the problem of finding marked vertices in a graph by a quantum walk. That is, we mark some set of vertices $|M|$, start at an arbitrary node in the graph and move according to the walk until we find the marked nodes. This explanation of the algorithm is based on de Wolf's lecture notes \cite{quantum_algorithm}, which builds on Santa's paper \cite{santa} that presents a slightly simplified version of the MNRS algorithm \cite{mnrs}. The basis states in the quantum walk search algorithm have two registers, one corresponding to the current node and the other corresponding to the previous node. That is, the basis states corresponds to the \textit{edges} in the graph. We denote the quantum walk based on the classical Markov chain with transition matrix $P$ by the unitary operation $W(P)$ on $\mathcal{H}$. We also define $\ket{p_x} = \sum_y \sqrt{P_{xy}}\ket{y}$ as the uniform superposition over the node $x$'s neighbors. Let $\ket{x}\ket{y}$ be a basis state. We define the basis state $\ket{x}\ket{y}$ as ``good'' if $x$ is a marked node. Otherwise, we refer to it as ``bad''. We now introduce ``good'' and ``bad'' states: 

\begin{equation}
    \ket{G} = \frac{1}{\sqrt{|M|}} \sum_{x \in M} \ket{x} \ket{p_x}, \;
    \ket{B} = \frac{1}{\sqrt{N-|M|}} \sum_{x \notin M} \ket{x} \ket{p_x},
\end{equation}

which are the superpositions over good and bad basis states. Next, let us define $\epsilon = |M|/N$ and $\theta = \arcsin(\sqrt{\epsilon})$. \\

In short, the algorithm consists of three steps: 

\begin{outline}[enumerate]
 \1 Set up the initial state $\ket{U} = \frac{1}{\sqrt{N}} \sum_{x} \ket{x} \ket{p_x} = \sin{\theta} \ket{G} + \cos{\theta} \ket{B}$, a uniform superposition over all edges
 \1 Repeat $O(1/\sqrt{\epsilon})$ times:
   \2 Reflect through $\ket{B}$
   \2 Reflect through $\ket{U}$
\1 Perform a measurement in the computational basis 
\end{outline}

Step $1$ is easily implemented with Hadamard gates. The reflection through $\ket{B}$ can be implemented with a phase oracle that shifts the phase of $x$ if $x$ is in the first register, and leaves the circuit unchanged otherwise. \\

Step 2(b) is equivalent to finding a unitary $R(P)$ that performs the following mapping: 
\begin{align}
\label{eq:mapping_1}
    \ket{U} &\mapsto \ket{U}, \: \text{and} \\
    \ket{\psi} &\mapsto -\ket{\psi}, \: \forall \ket{\psi} \text{in the span of eigenvectors of $W(P)$ that are orthogonal to $\ket{U}$}
\label{eq:mapping_2}
\end{align}

To find this operator we apply phase estimation on $W(P)$.
Above we defined $W(P)$ as the evolution operator for the random walk. As we saw in section \ref{sec:q_walk_coins}, this is a unitary operator. From this follows that the eigenvalues of $W(P)$ have norm $1$. Because of this, we can write the eigenvalues of $W(P)$ on the form $e^{\pm 2i\theta_j}$. The unitary $W(P)$ has one eigenvector with corresponding eigenvalue $1$, which is $\ket{U}$. This is given by $\theta_1=0$. $R(P)$ will find this vector $\ket{U}$ by adding a register with ancilla qubits and perform phase estimation with precision $O(1/\sqrt{\delta})$, where $\delta$ is the spectral gap of $P$. To do this, we need to apply $W(P)$ $O(1/\sqrt{\delta})$ times. Let $\ket{w}$ be an eigenvector of $W(P)$ with eigenvalue $e^{\pm 2i\theta_j}$ and assume that $\Tilde{\theta_j}$ is the best approximation to $\theta_j$ given by the phase estimation. The operation $R(P)$ that performs the mappings in \eqref{eq:mapping_1} and \eqref{eq:mapping_2} for $\ket{w}$ is given by \cite{quantum_algorithm} \\

\begin{equation}
    \ket{w} \ket{0} \mapsto \ket{w} \ket{\Tilde{\theta_j}} \mapsto (-1)^{|\Tilde{\theta_j} \neq 0|} \ket{w} \ket{\Tilde{\theta_j}} \mapsto (-1)^{|\Tilde{\theta_j} \neq 0|} \ket{w} \ket{0}.
\end{equation}

The hitting time is the number of iterations of step 2 required to maximize the probability that the circuit collapses to one of the marked states, and is a central part in this project. We will compare this theoretical value, $O(1/\sqrt{\epsilon})$, to the results we get when we run the algorithm for a few different graphs.

\section{Method} \label{sec:method}

\subsection{Implementation of Grover coin}
\begin{figure}[h!]
\centering
\includegraphics[width=5cm]{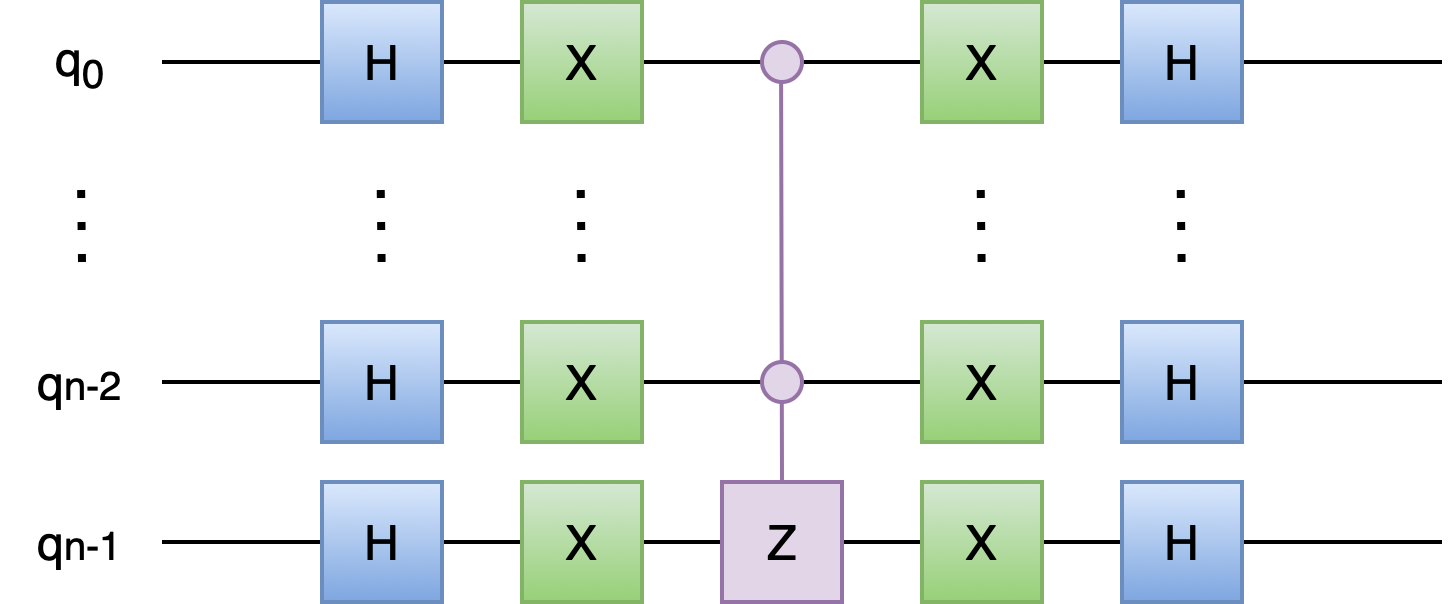}
\caption{Implementation an of $n$-qubit Grover coin.}
\label{fig:grover_coin}
\end{figure}

The Grover coin is a crucial part of many of the walks we present below. The implementation consists of Hadamard- and NOT gates along with a controlled-Z gate. Figure \ref{fig:grover_coin} shows the implementation of a Grover coin of an arbitrary size $n$. When we henceforth discuss the Grover coin implementation, this is what we refer to. 

\subsection{Coined quantum walks}
\subsubsection{Hypercube} \label{sec:hypercube}
\begin{figure}[H] 
\centering
\includegraphics[width=8cm]{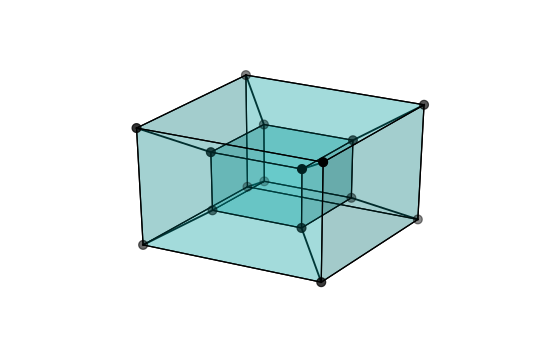}
\caption{A 4-dimensional hypercube graph with 16 nodes.}
\label{fig:hypercube_graph}
\end{figure}

A hypercube is the $n$-dimensional counterpart of the $3$-dimensional cube. All nodes have degree $n$, and the hypercube has a total of $N=2^n$ nodes. A $4$-dimensional hypercube is shown in Figure \ref{fig:hypercube_graph}. \\


We can represent the nodes in a hypercube graph by $n$-tuples of binary numbers. The binary representation of the neighbors of a node will differ by only one binary number. For example, in the 4-dimensional hypercube, the neighbors to $0000$ are $0001$, $0010$, $0100$, and $1000$. Thus a node is connected to all nodes to which the Hamming distance is 1 \cite{qwalks}. The edges are also labeled. Two neighboring nodes that differ in the \textit{a}:th bit are connected by the edge labeled $a$. \\

The Hilbert space representing a coined quantum walk on the hypercube is $\mathcal{H} = \mathcal{H}^n \otimes \mathcal{H}^{2^n}$, where $\mathcal{H}^n$ denotes the coin space and $\mathcal{H}^{2^n}$ the walker's position. The computational basis is

\begin{equation}
    \big\{ \ket{a,\vec{v}}, 0 \leq a \leq n-1, \vec{v} \in  \{(00...00), (00...01), ....., (11...11 )\} \big\}. 
\end{equation}

The value of the coin computational basis $a$, which is associated with edge $a$, decides the direction to move. If the $a=0$, the walker will go to the node where the first binary value differs from the current node. If $a=1$, the walker will move to the node in which the second value differs from the current, et cetera.  Let $\vec{e}_a$ be an n-tuple where all binary values, except the value with index $a$, are $0$. Then the shift operator $S$ moves the walker from the state $\ket{a} \ket{\vec{v}}$ to $\ket{a} \ket{\vec{v} \oplus \vec{e}_a}$ \cite{qwalks}:

\begin{equation}
    S \ket{a} \ket{\vec{v}} = \ket{a} \ket{\vec{v} \oplus \vec{e}_a}.
\end{equation}

We use the Grover coin, $G$, for this walk, and thus the evolution operator is

\begin{equation}
    U = SG.
\end{equation}

\begin{figure}[H]
\centering
\includegraphics[width=13cm]{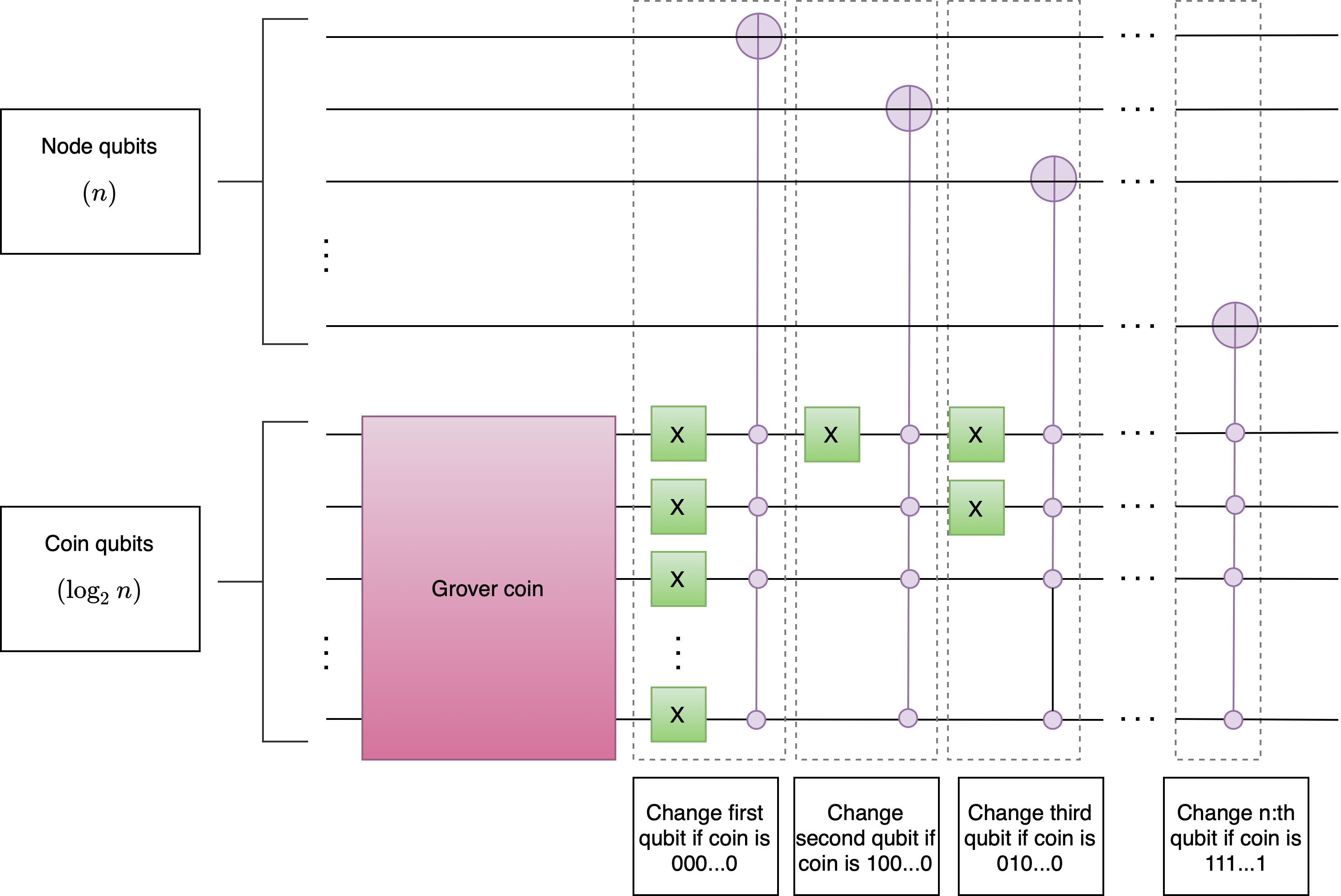}
\caption{Implementation of one step of the quantum walk with Grover coin on an $n$-dimensional hypercube. $2^n$ nodes.}
\label{fig:hypercube_implementation}
\end{figure}

The implementation of the quantum walk on an $n$-dimensional hypercube is shown in Figure \ref{fig:hypercube_implementation}.

\subsubsection{2-dimensional lattice with boundary conditions} \label{sec:2d_lattice}
\begin{figure}[H]
\centering
\includegraphics[width=10cm]{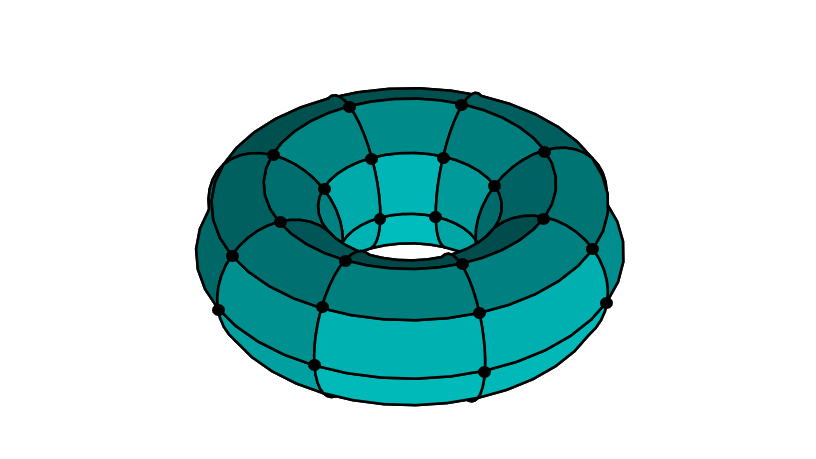}
\caption{2-dimensional lattice with periodic boundary
conditions. $8 \times 8$ nodes.}
\label{fig:2d_lattice_graph}
\end{figure}

A 2-dimensional lattice graph with periodic boundary conditions has the form of a torus as we see in Figure \ref{fig:2d_lattice_graph}. \\

The 2-dimensional lattice graph has a total of $\sqrt{N} \times \sqrt{N}$ nodes, and the walker can move in either the $x$-direction or the $y$-direction. By taking $\sqrt{N}$ steps in $x-$ or $y$-direction, the walker ends up in its initial position. We define the coin space by $\ket{d,s}$, where $d$ decides in what direction, $x$ or $y$, the walker should move, and $s$ decides whether it should increase or decrease. The position space is represented by $\ket{x,y}$ \cite{qwalks}. The computational basis for the entire quantum walk is
 \begin{equation}
      \{\ket{d,s,x,y}, d,s \in \{0,1\}, 0 \leq x,y \leq  \sqrt{N} -1  \}.
 \end{equation}
 
There is also a shift operator, $S$, that operates as follows: 
 \begin{equation}
S \ket{d,s} \ket{x,y} = \ket{d,s \oplus 1} \ket{x + (-1)^s \delta_{d0} , y + (-1)^s \delta_{d1}}.
 \end{equation}

If $d=0$ and $s=0$, $y$ is left unchanged and $x$ is increased by one after the application of $S$. If $x$ is changed, $y$ is left unchanged, and vice versa. We can also see that the shift operator changes the coin state so that $s$ gets inverted: $\ket{d,s \oplus 1}$. This increases the speed for searching algorithms for this graph \cite{qwalks}. We implement the walk with a Grover coin, so the evolution operator is
 \begin{equation}
     U = SG. 
 \end{equation}
 
\begin{figure}[H]
\centering
\includegraphics[width=15cm]{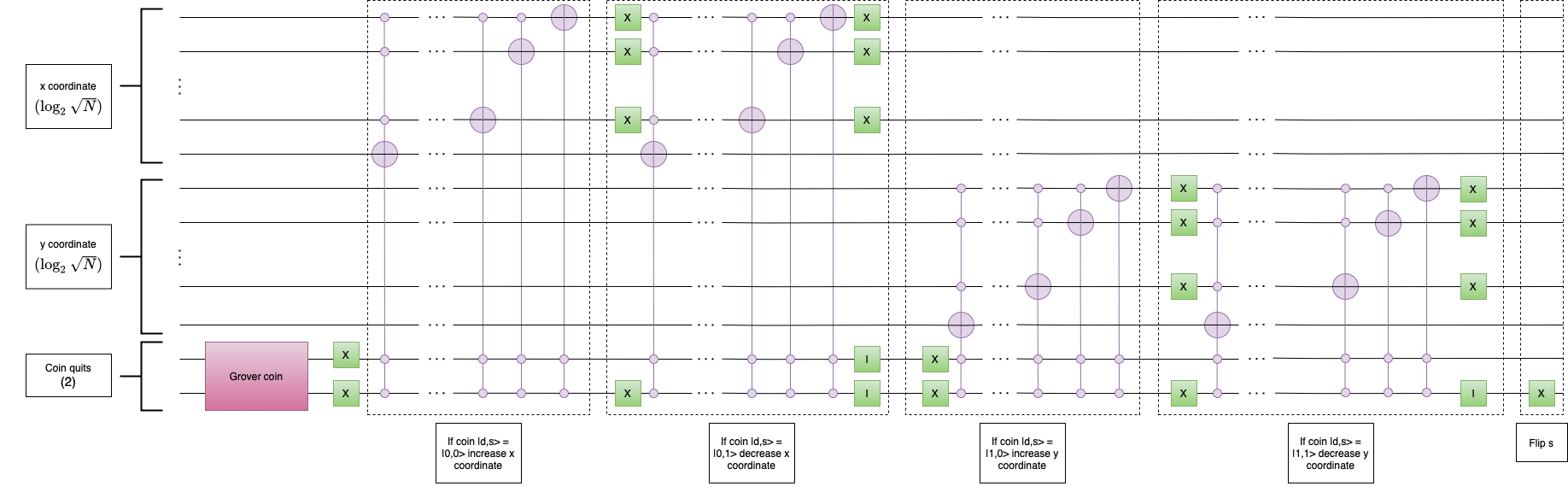}
\caption{2-dimensional lattice implementation for one step with Grover coin. $\sqrt{N} \times \sqrt{N}$ nodes.}
\label{fig:torus_implementation}
\end{figure}

 Figure \ref{fig:torus_implementation} shows the implementation of a quantum walk on a 2-dimensional $\sqrt{N} \times \sqrt{N}$ graph. 

\subsubsection{Complete bipartite graph} \label{sec:complete_bipartite}

\begin{figure}[H]
\centering
\includegraphics[width=8cm]{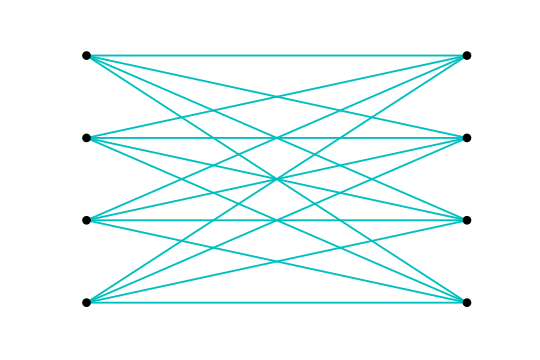}
\caption{Complete bipartite graph with 8 nodes.}
\label{fig:bipartite_graph}
\end{figure}

Figure \ref{fig:bipartite_graph} shows a complete bipartite graph with a total of 8 nodes, 4 in each set. We implement the quantum walk on bipartite graphs with the same number of nodes in both sets. There is a total of $N$ nodes, and each node is connected to $\frac{N}{2}$ neighbors. The computational basis for this walk is
\begin{equation}
    \{\ket{c,j}, 0 \leq c \leq \frac{N}{2} - 1,  0 \leq j \leq N -1  \}.
\end{equation}

\begin{figure}[H]
\centering
\includegraphics[width=10cm]{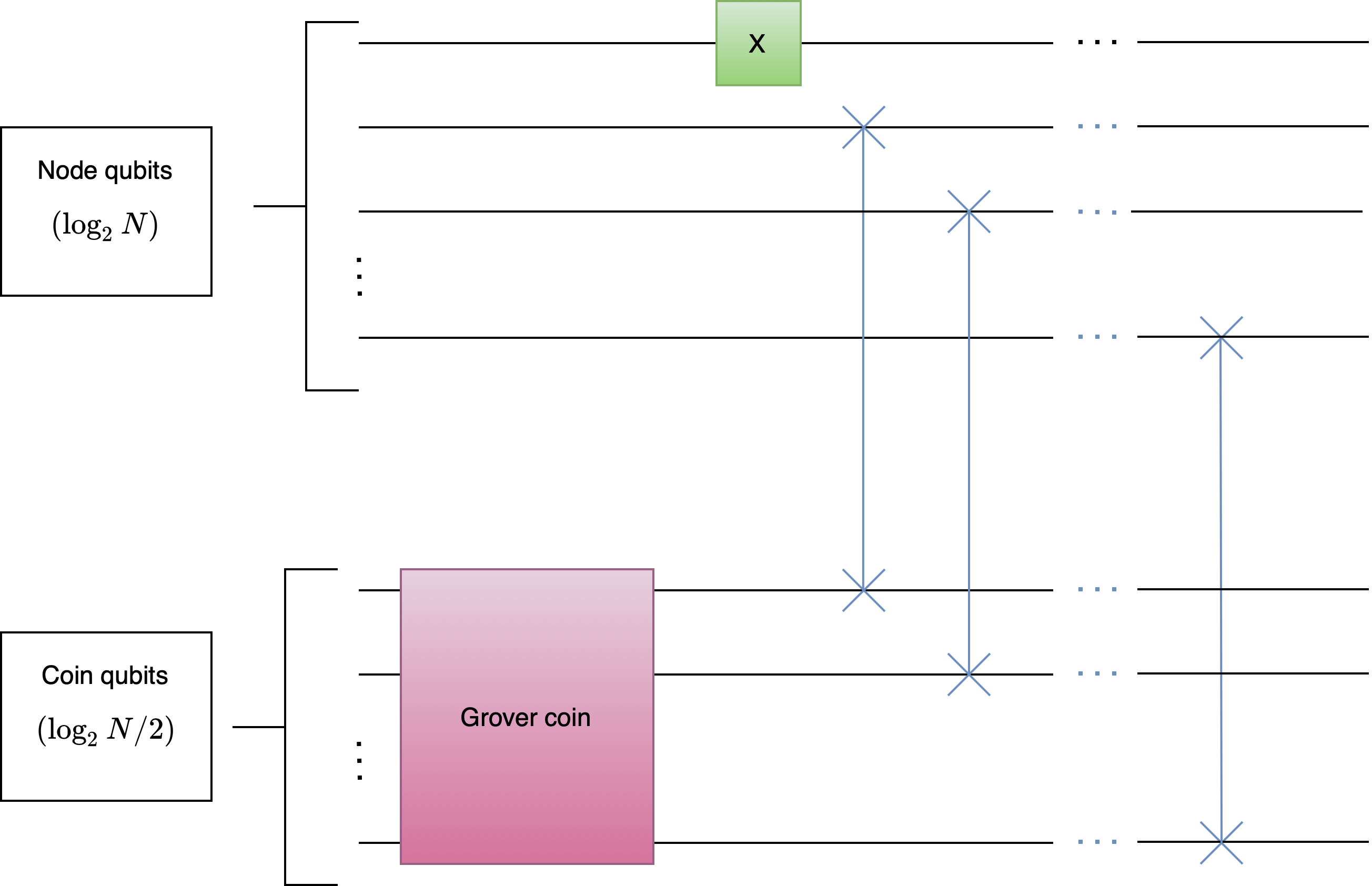}
\caption{Implementation of quantum walk on complete bipartite graph with $N$ nodes.}
\label{fig:impl_bipartite_graph}
\end{figure}

The nodes in the graph are numbered such that all nodes belonging to the same set will start with the same binary number. For example, in the complete bipartite graph in Figure \ref{fig:bipartite_graph}, the nodes on the left will have the binary representations $000$, $001$, $010$, $011$ and belong to the same set. Similarly, the nodes to the right belong to the same set and have the labels $100$, $101$, $110$, $111$. If $2^n = N$ and $j_0 j_1 .... j_{n-1}$ is the binary representation of $j$, where $j = 2^{n-1} j_0 + 2^{n-2} j_1 + .... + 2^0 j_n$, and similarly $c_0 c_1 .. c_{n-2}$ is the binary representation of $c$, then the shift operator will act in the following way:
\begin{equation}
    S \ket{c} \ket{j} = \ket{j_1 j_2 .... j_{n-1}} \ket{(j_0 \oplus 1) c_0 c_1 ... c_{n-2} }.
\end{equation}
The shift operator swaps the coin space with the position space for all qubits except for the position qubit $j_0$, which flips. Figure \ref{fig:impl_bipartite_graph} shows an implementation of the quantum walk on a complete bipartite graph with $N$ nodes. We implement the walk with a Grover coin, so the evolution operator is $U=SG$.

\subsubsection{Complete graph} \label{sec:compl_graph}
\begin{figure}[H]
\centering
\includegraphics[width=8cm]{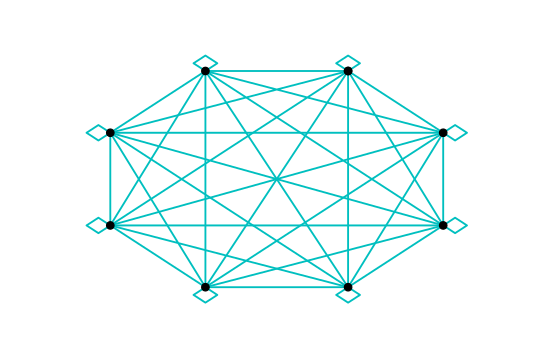}
\caption{Complete graph with 8 nodes.}
\label{fig:complete_graph}
\end{figure}

Here we present the implementation of a coined quantum walk on a complete graph with self loops. Figure \ref{fig:complete_graph} shows a complete graph with 8 nodes. In a complete graph with $N$ nodes, every node has $N$ neighbors, itself included. The computational basis for the walk is
\begin{equation}
    \{\ket{c,j}, 0 \leq c \leq N - 1,  0 \leq j \leq N - 1  \}.
\end{equation}

The shift operator swaps the coin qubit values with the position qubit values, so it acts in the following way:
\begin{equation}
        S \ket{c} \ket{j} = \ket{j} \ket{c}.
\end{equation}

\begin{figure}[H]
\centering
\includegraphics[width=10cm]{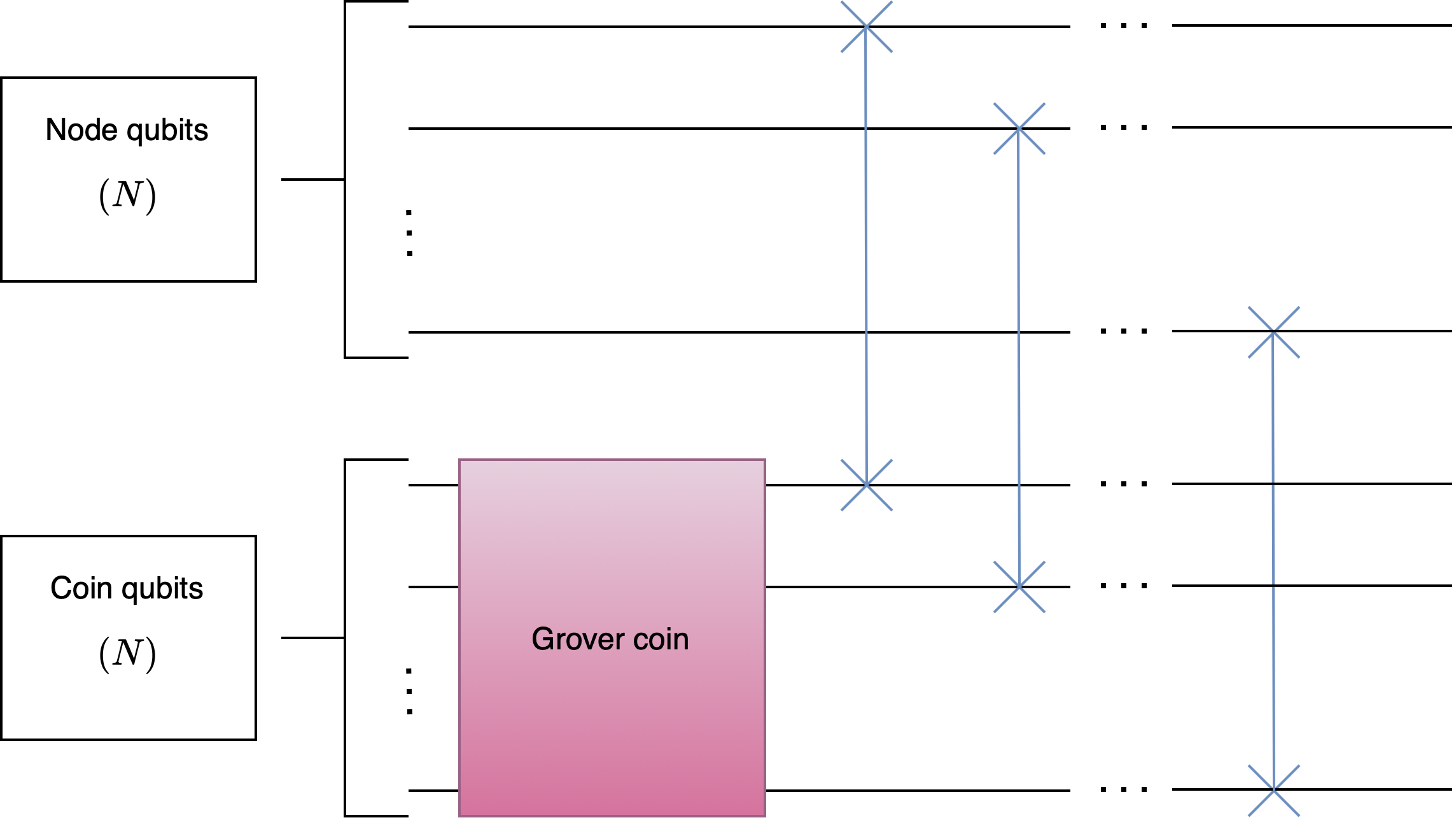}
\caption{Implementation of quantum walk on complete graph with $N$ nodes.}
\label{fig:impl_complete_graph}
\end{figure}

Figure \ref{fig:impl_complete_graph} shows an implementation of the walk on a complete graph with $N$ nodes using a Grover coin. The evolution operator for the walk is $U=SG$.

\subsection{Implementation of quantum walk search algorithm} \label{sec:qwalk_implementation}
This quantum walk search algorithm, which we introduced in section \ref{section:qwa}, makes it possible to find a marked set of nodes in $O(1/\sqrt{\epsilon})$ steps, $\epsilon =  |M|/N$, where $M$ is the set of marked nodes and $N$ is the total number of nodes. This algorithm is originally used with Szegedy's quantum walk, where two node registers are used to represent the quantum state. However, a coined walk with Grover coin is equivalent to Szegedy's quantum walk, and its implementation is usually less complicated. Therefore, we chose to implement the algorithm with coined walks. \\

\begin{figure}[H] 
\centering
\includegraphics[width=15cm]{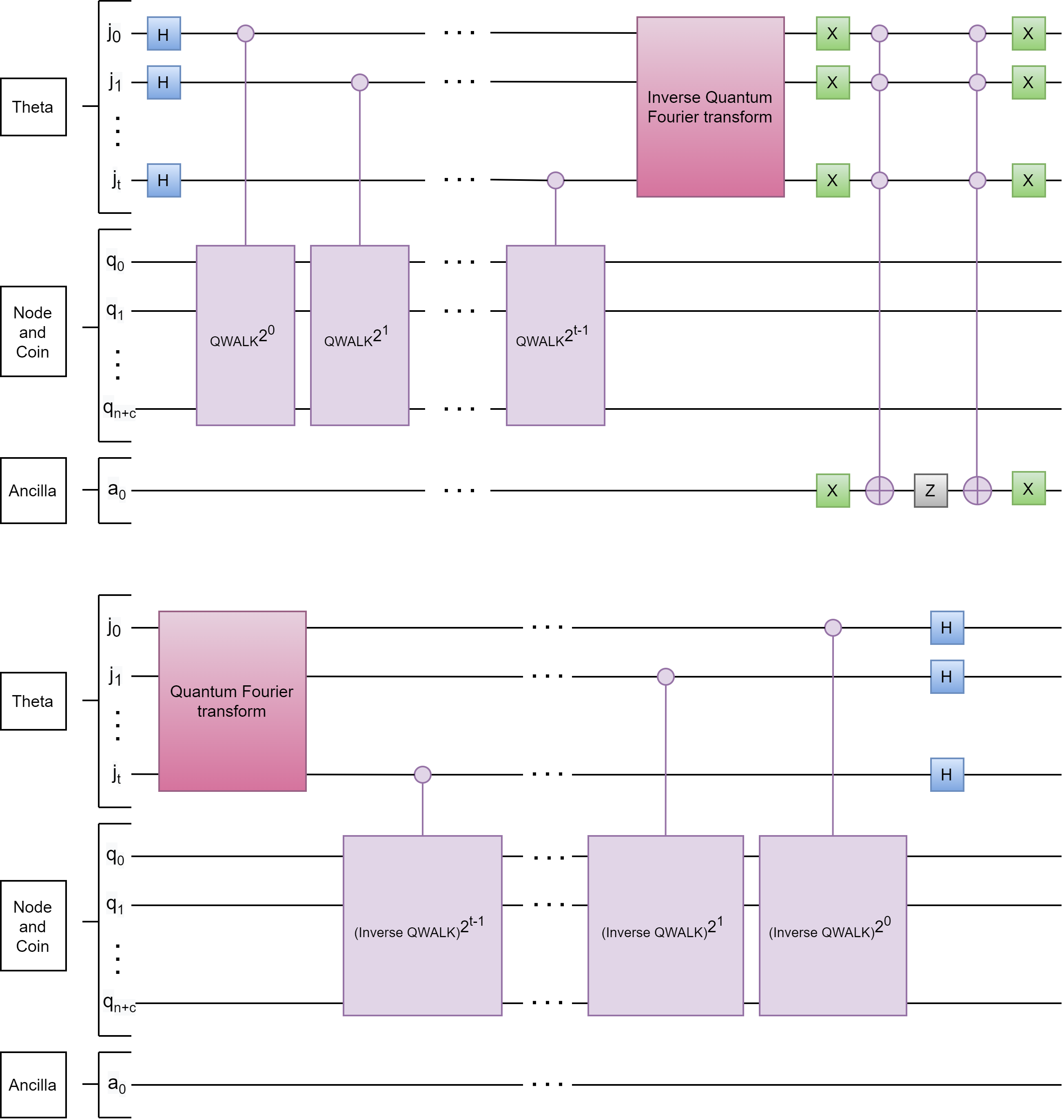}
\caption{Step 2(b) of the quantum walk search algorithm. A phase estimation followed by marking an ancilla qubit if $\theta \neq 0$ and a reversed phase estimation.}
\label{fig:qwalk_alg_1}
\end{figure}

\begin{figure}[H] 
\centering
\includegraphics[width=15cm]{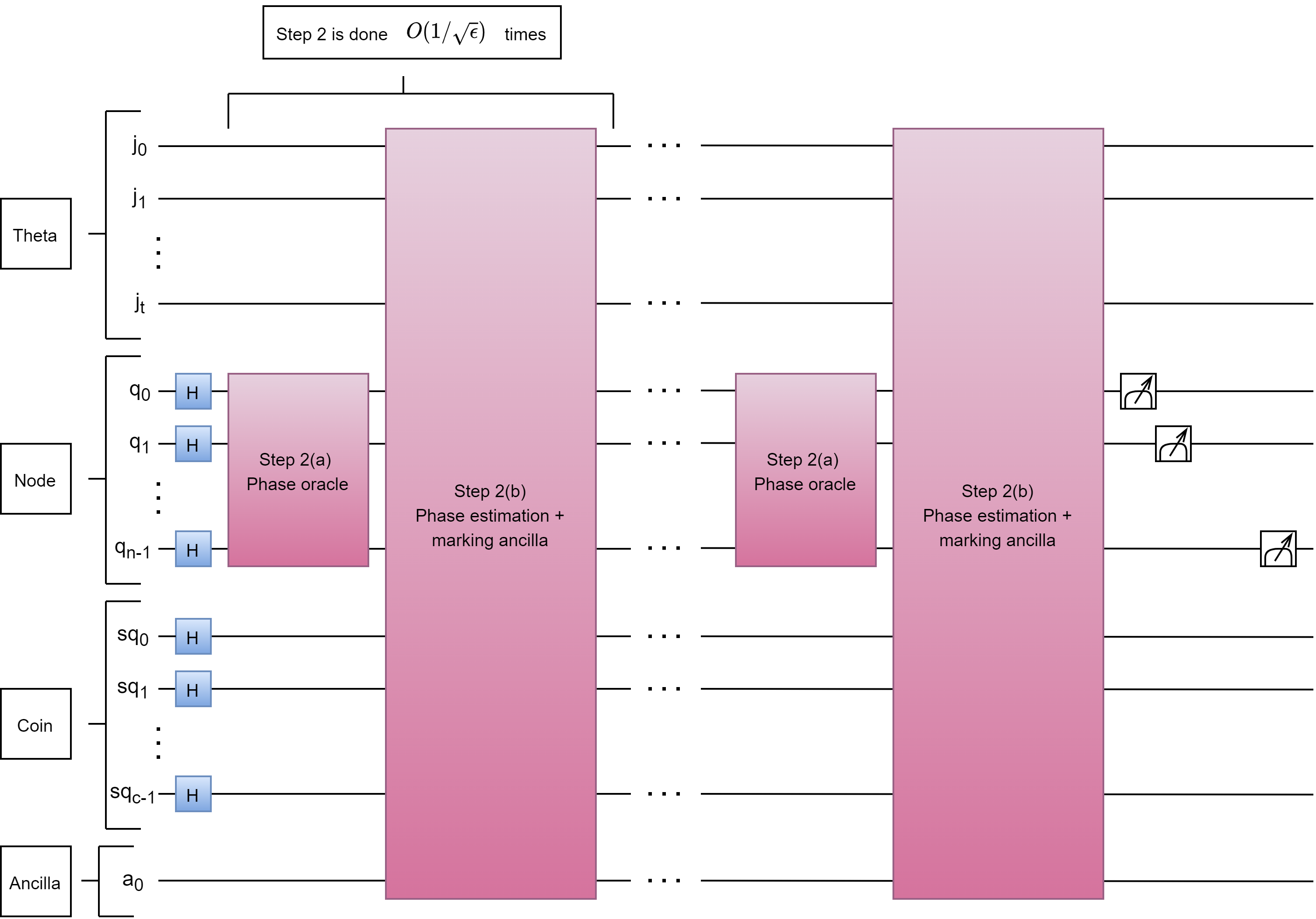}
\caption{Complete implementation of quantum walk search algorithm.}
\label{fig:qwalk_final}
\end{figure}

We achieve step 1, a uniform superposition over all edges, by applying Hadamard gates to the node qubits and the coin qubits. For step 2(a), we implement a phase oracle. The implementation of step 2(b) is shown in Figure \ref{fig:qwalk_alg_1}. This step consists of a phase estimation followed by marking all quantum states where $\theta \neq 0$. We do this by rotating an ancilla qubit. In the last part of this step, we reverse the phase estimation. The QWALK gate in this implementation is one step of a quantum walk, which is described in the previous sections (\ref{sec:hypercube}, \ref{sec:2d_lattice}, \ref{sec:complete_bipartite}, \ref{sec:compl_graph}). The number of theta qubits depends on the precision of $\theta$.  Figure \ref{fig:qwalk_final} shows the complete quantum walk search algorithm. 

\section{Result} \label{sec:result}
This section presents the experimental results of the quantum hitting times for 4-dimensional hypercube, 2-dimensional lattice, complete bipartite graph, and complete graph. We have implemented the quantum walk search algorithm as we showed in section \ref{sec:qwalk_implementation} in Qiskit on a noise-free simulator and a simulator with the noise model of the Melbourne computer. The state vector plots show the state vectors in iterations 1-4 in the noise-free simulator. We also show the measurement outputs for both the noise-free and the noisy simulator and compare their results.

\subsection{4-dimensional hypercube}
\begin{figure}[H]
\centering
\includegraphics[width=15cm]{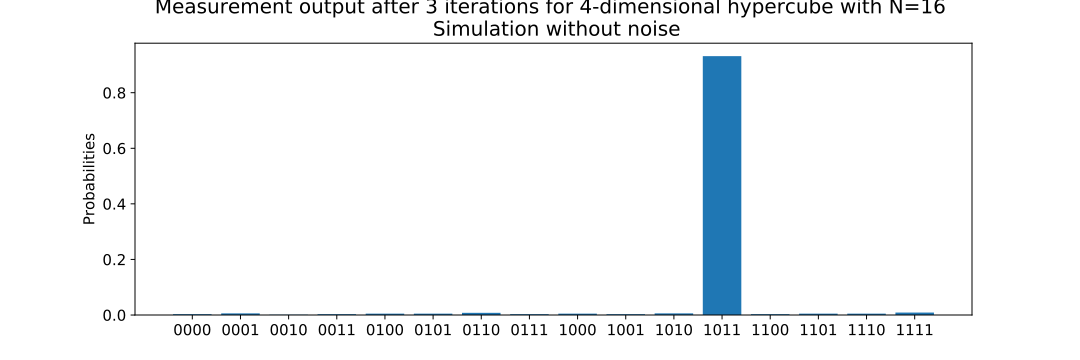}
\caption{Measurement of quantum walk search algorithm implemented on a noise-free simulator for a 4-dimensional hypercube after 3 iterations with node 1011 marked. The circuit collapses to the marked node $93.2\%$ of the times.  }
\label{fig:res_4d_hit}
\end{figure}

Figure \ref{fig:4d_hit} below shows the state vector simulations for the 4-dimensional hypercube with 16 nodes in the first four iterations on a noise-free simulator in Qiskit. Initially, the states are in superposition, and we have marked state $1011$. We see that the probability that the circuit collapses to state $1011$ increases until the third iteration, where it reaches the maximum probability before it starts to decrease again. Thus, the hitting time for the graph is $3$. Figure \ref{fig:res_4d_hit} shows the result from running the full algorithm on a noise-free simulator, followed by a measurement after 3 iterations. As we see in the figure, the circuit collapses to the marked state a significant majority of the times, over $93\%$. \\

\begin{figure}[H]
\centering
\hspace*{-0.7in}
\includegraphics[width=19cm]{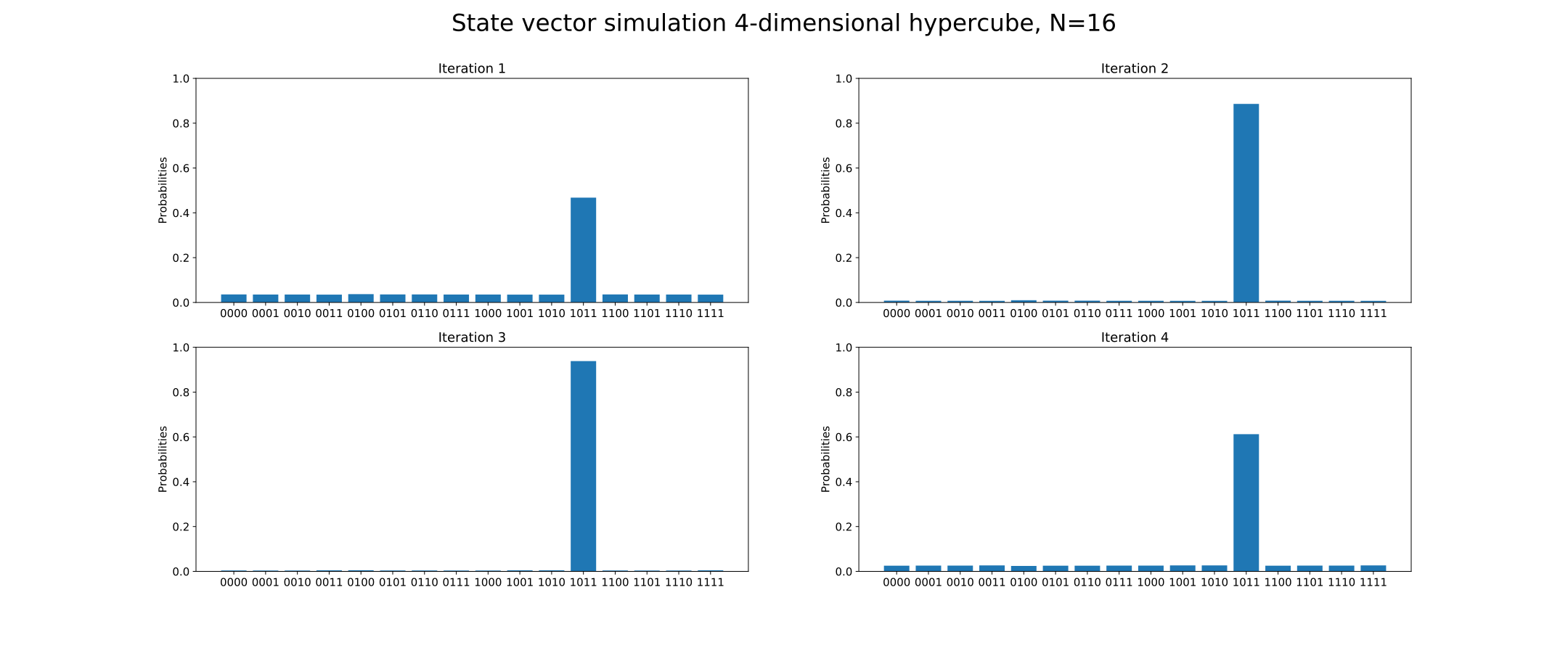}
\caption{State vectors for the quantum walk search algorithm on a 4-dimensional hypercube implemented on a noise-free simulator. 1011 is marked. }
\label{fig:4d_hit}
\end{figure}

\begin{figure}[H]
\centering
\includegraphics[width=15cm]{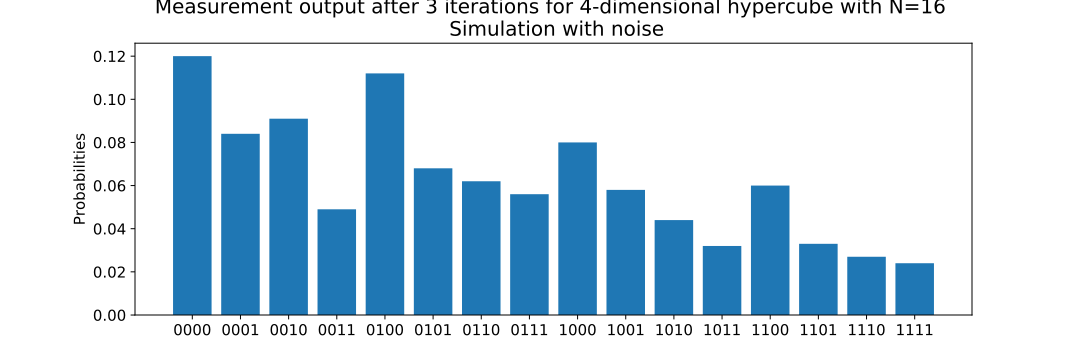}
\caption{Measurement of quantum walk search algorithm for a 4-dimensional hypercube graph with 16 nodes after 3 iterations with node 1011 marked. The circuit is implemented on the fake Melbourne simulator that simulates the noise of the real Melbourne quantum computer. The circuit collapses to the marked node $3.2\%$ of the times. }
\label{fig:hyp_noise}
\end{figure}

Figure \ref{fig:hyp_noise} shows the measurement output after running the same program on a noisy quantum simulator. Again, we have marked node 1011. As we see in the figure, the circuit does not collapse to the marked node most of the time; it is not even the node with the highest output probability. In the noisy simulation, the marked node is the output only $3.2\%$ of the times, compared to the noise-free circuit where the circuit collapsed to the marked node $93.2\%$ of the times. These results illustrate the extent to which noise influences the behavior of the algorithm.

\subsection{2-dimensional lattice}
\begin{figure}[H]
\centering
\hspace*{-0.7in}
\includegraphics[width=19cm]{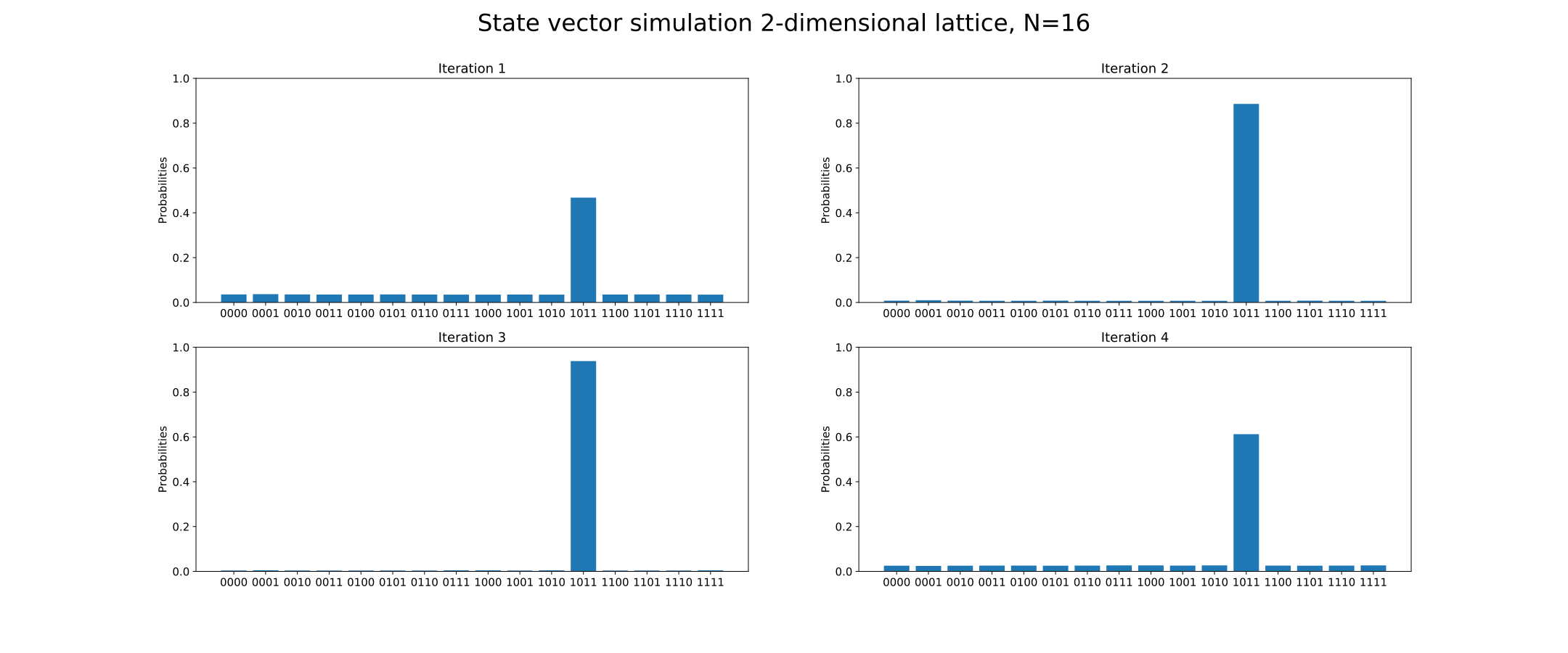}
\caption{Noise-free state vectors for the 2-dimensional lattice. 1011 is marked. }
\label{fig:2d_hit}
\end{figure}

\begin{figure}[H]
\centering
\includegraphics[width=15cm]{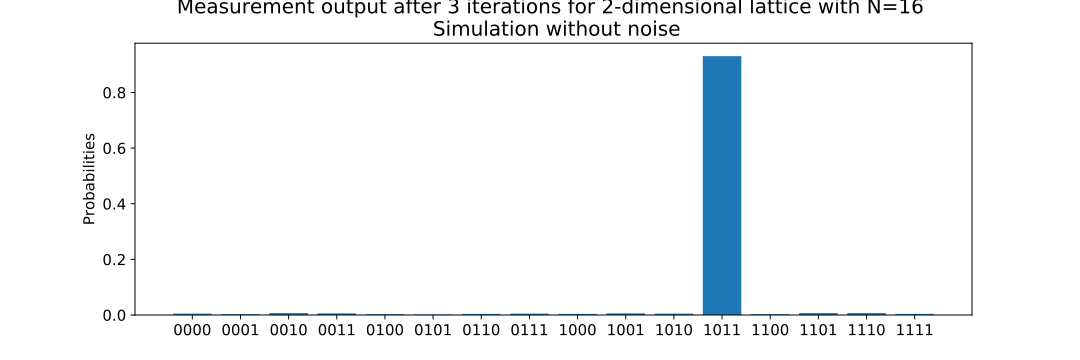}
\caption{Measurement of quantum walk search algorithm for a 2-dimensional lattice after three iterations with node 1011 marked. The circuit collapses to the marked node $93.1\%$ of the times and the program was executed on a simulator without noise. }
\label{fig:res_toroid}
\end{figure}

For the 2-dimensional lattice with 16 nodes and one marked node ($1011$), the hitting time is 3. We can see this in the state vector plots from the noise-free simulator presented in Figure \ref{fig:2d_hit}, where the maximum probability that the circuit collapse to $1011$ occurs in the third iteration. Figure \ref{fig:res_toroid} shows the distribution of the result after 3 iterations of the algorithm for 1024 executions without noise. In $93.1\%$ of the runs, the circuit collapses to the marked state. \\

\begin{figure}[H]
\centering
\includegraphics[width=15cm]{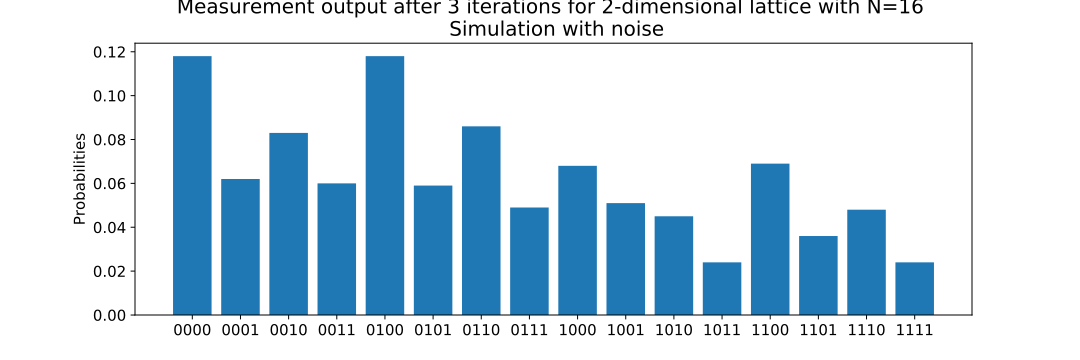}
\caption{Measurement of quantum walk search algorithm for a 2-dimensional lattice graph with 16 nodes after three iterations with node 1011 marked. This simulation was done on the fake Melbourne simulator to simulate a real quantum computer's noise when running this circuit. The circuit collapses to the marked node $2.4\%$ of the times.  }
\label{fig:tor_noise}
\end{figure}

Figure \ref{fig:tor_noise} shows the output of same implementation executed on the fake Melbourne simulator. Here, the circuit collapses to the marked node the least number of times. 

\subsection{Complete bipartite graph}
\begin{figure}[H]
\centering
\hspace*{-0.7in}
\includegraphics[width=19cm]{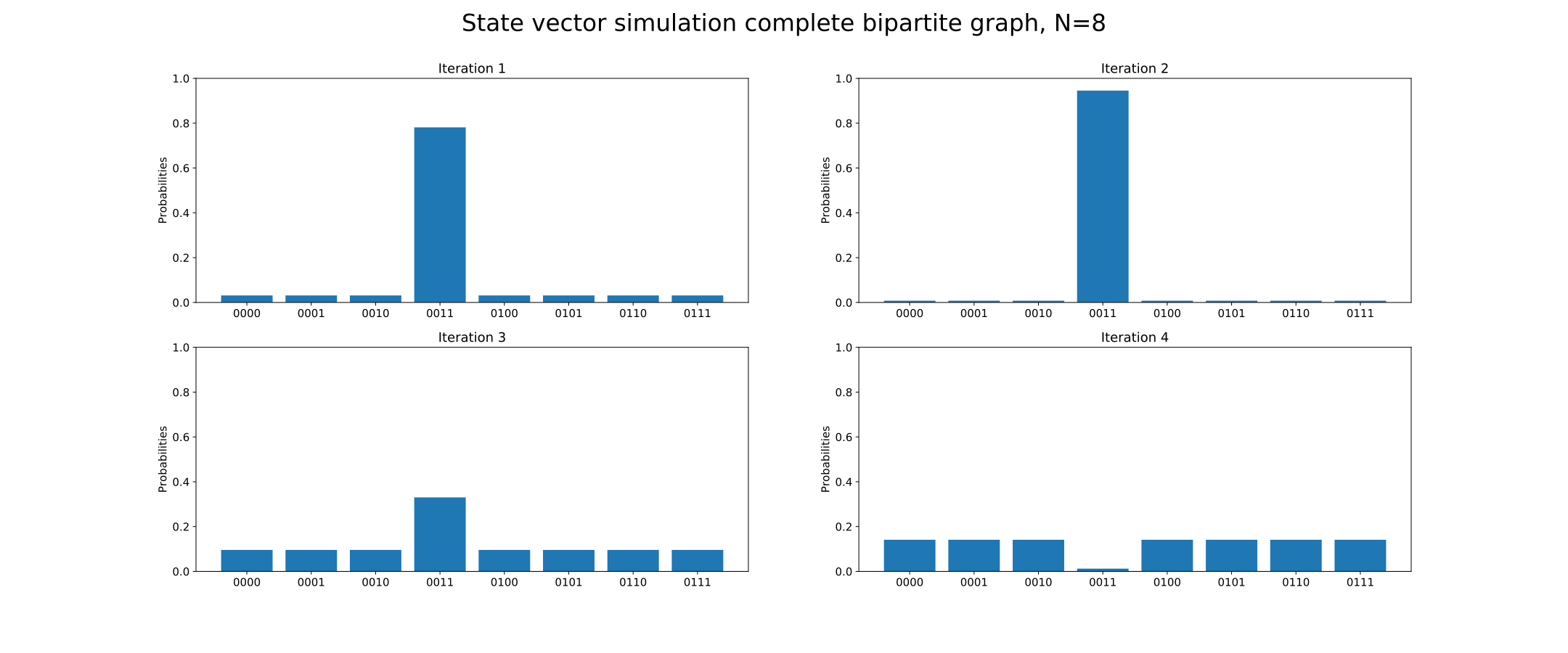}
\caption{Quantum walk search algorithm state vector for a complete bipartite graph with node 011 marked. Executed on noise-free simulator.}
\label{fig:cpl_bipartite}
\end{figure}

When running the quantum walk search algorithm for the complete bipartite graph, we marked state $011$. Figure \ref{fig:cpl_bipartite} shows the noise-free state vectors for the first four iterations. As we see in the figure, the circuit collapses to the marked state with the highest probability after the second iteration, and we conclude that the hitting time is 2. In Figure \ref{fig:cpl_bipartite}, we have run the algorithm on the same simulator with two iterations 1024 times, and it collapses to state $011$ $94.5\%$ of the times. \\

\begin{figure}[H]
\centering
\includegraphics[width=15cm]{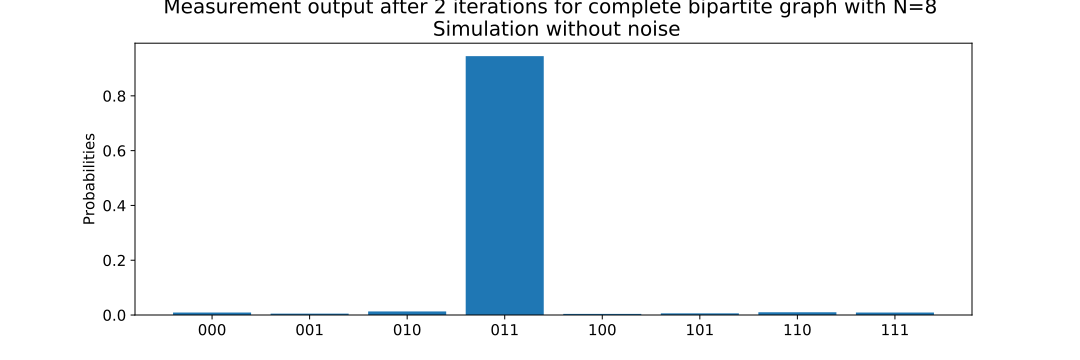}
\caption{Measurement of quantum walk search algorithm for a complete bipartite graph with 8 nodes after 2 iterations on noise-free simulator with node 011 marked. The circuit collapses to the marked node $94.5\%$ of the times. }
\label{fig:res_cpl_bip}
\end{figure}

\begin{figure}[H]
\centering
\includegraphics[width=15cm]{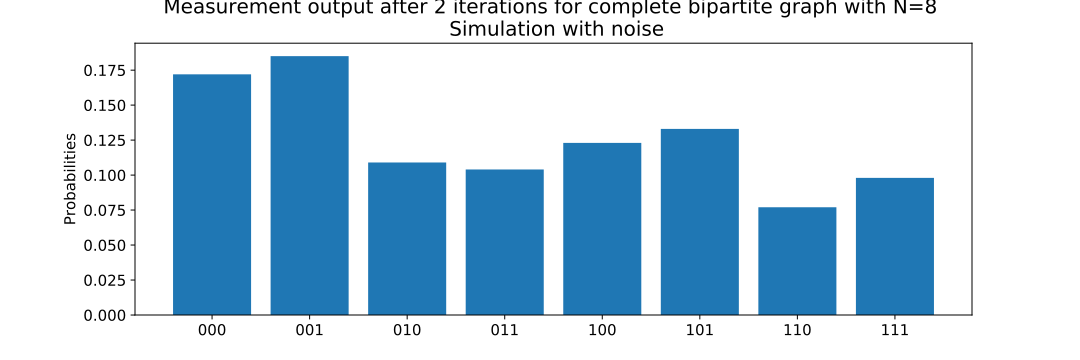}
\caption{Measurement of quantum walk search algorithm for a complete bipartite graph with 8 nodes after 2 iterations with node 011 marked on fake Melbourne simulator. The circuit collapses to the marked node $10.4\%$ of the times. }
\label{fig:bip_noise}
\end{figure}

We also implement the algorithm on a simulator with the fake Melbourne backend. The measurement output after 2 iterations after marking node 011 is shown in Figure \ref{fig:bip_noise}. Compared to the noise-free simulator, where the circuit collapsed to the marked node $94.5\%$ of the times, the same percentage is only $10.4\%$ on the noisy device. 

\subsection{Complete graph}
\begin{figure}[H]
\centering
\hspace*{-0.7in}
\includegraphics[width=19cm]{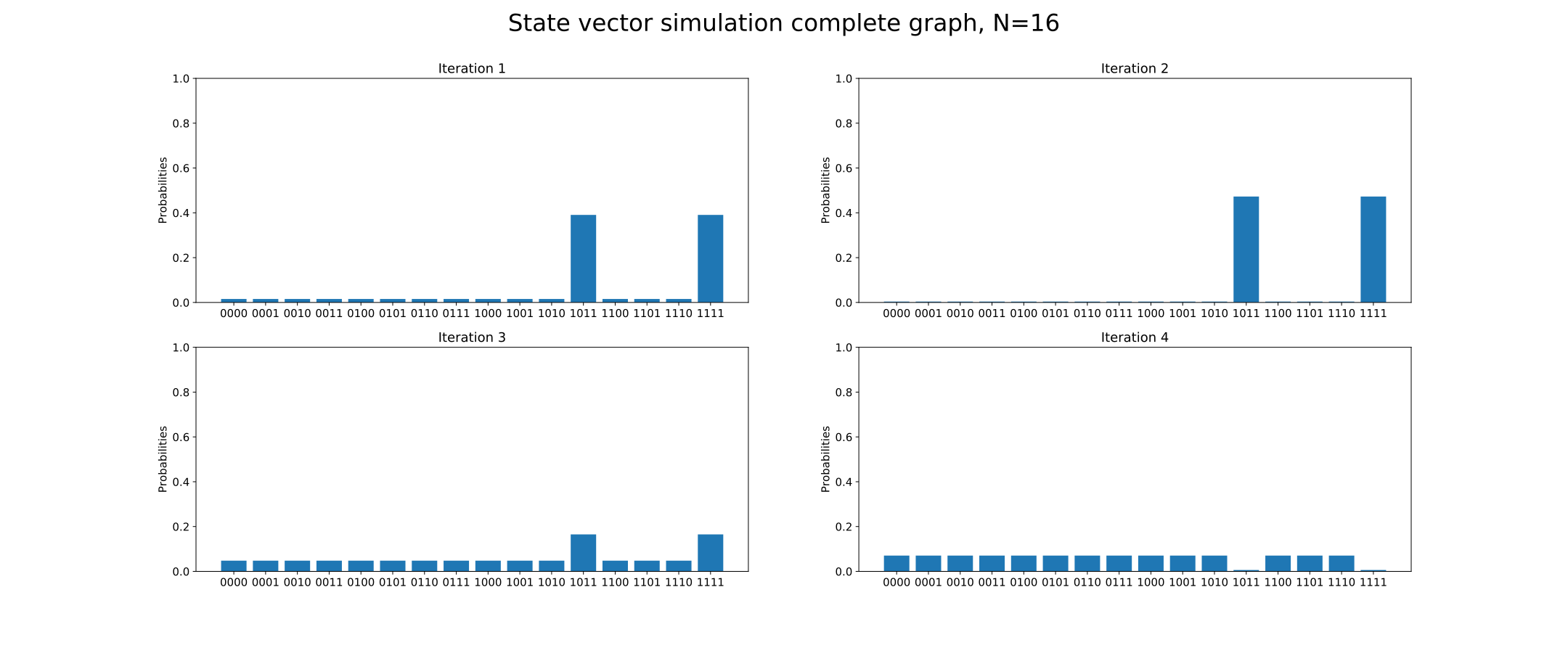}
\caption{Noise-free state vector simulations for a complete graph with 16 nodes. 1011 and 1111 is marked.}
\label{fig:cpl_graph}
\end{figure}

In Figure \ref{fig:cpl_graph} we see the state vectors for the complete graph with 8 nodes and two marked nodes, $1011$ and $1111$. With the same logic as above, we conclude that the hitting time is 2. \\

\begin{figure}[H]
\centering
\includegraphics[width=15cm]{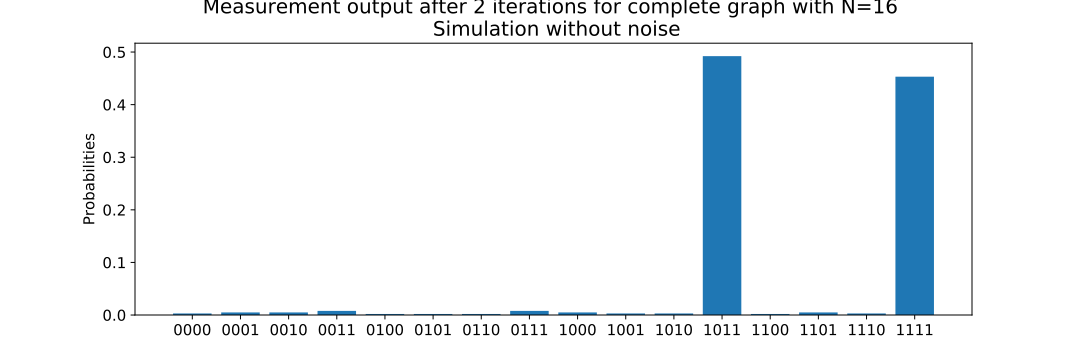}
\caption{Measurement of quantum walk search algorithm for a complete graph with 16 nodes after 2 iterations with two nodes, 1011 and 1111, marked. The circuit collapses to a marked node $94.5\%$ of the times. }
\label{fig:res_compl}
\end{figure}

\begin{figure}[H]
\centering
\includegraphics[width=15cm]{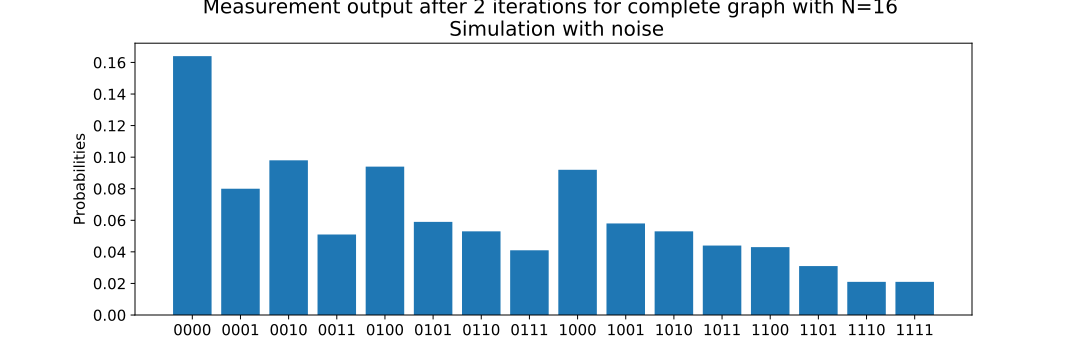}
\caption{Measurement of the quantum walk search algorithm for a complete graph with 16 nodes after 2 iterations with nodes 1011 and 1111 marked. This simulation was done on the fake Melbourne simulator to simulate the noise a real quantum computer would have when running this circuit. The circuit collapses to 1011 and 1111 $4.4\%$ and $2.1\%$ of the times respectively. }
\label{fig:cpl_noise}
\end{figure}

We see the final, noise-free output of the algorithm after two iterations in Figure \ref{fig:res_compl}. The circuit collapses to state $1011$ $49.2\%$ of the times and to state $1111$ $45.3\%$ of the times. That means that we end up in a marked state in $94.5\%$ of the executions. Figure \ref{fig:cpl_noise} shows the output from a noisy simulation. Here, the circuit collapses to one of the marked nodes $6.5\%$ of the times in total.

\section{Discussion} \label{discussion}
The number of iterations we needed to find the marked nodes for both the state vector simulations and the noise-free quantum computer simulations matches the theoretical values, which, for all of the graphs, is $O(1/\sqrt{\epsilon})$. $\epsilon$ is defined as $\epsilon = |M|/N$, where $M$ is the set of marked nodes and $N$ the size of the graph. For the simulations of complete- and complete bipartite graphs, this theoretical number was $O(1/\sqrt{\epsilon}) \approx O(2.8)$. In the simulations without noise, we could see that we needed two iterations to maximize the probability of returning the marked nodes for both of these graphs. This coincides well with the theoretical number. For the 4-dimensional hypercube and the 2-dimensional lattice, the theoretical number was $O(1/\sqrt{\epsilon})=O(4)$, and both of these graphs needed three iterations to find the marked nodes. As expected, the complete graph with two marked nodes and the complete bipartite graph with half the total nodes compared to all the other graphs needed a smaller number of iterations than the two other graphs, which had one marked node and a total of 16 nodes. When increasing the number of marked nodes or decreasing the number of total nodes, we expected the number of iterations to decrease, which our result showed. However, these results are from a noise-free simulator, which does not reflect the result we would get on an actual quantum device. 
\\

To understand how the quantum walk algorithm implementation would run on actual quantum computers, we also used simulations with noise. These results differed vastly from both the theoretical number of iterations and the results from the ideal, noise-free quantum computer. The output from simulating with noise did not have a high probability of returning the marked nodes; for most of the graphs, the percentage of times it found the marked nodes were much smaller than in the simulation without noise. Not only was it smaller than in the noise-free simulation, but it would also find an unmarked node most of the time. As discussed in section \ref{sec:noiseee}, the error rate from just using a single gate is quite large. When using as many gates as required in the quantum walk search algorithm, this error will be so immense that the output is essentially worthless. The large number of gates is not only due to the size of the algorithm, but also because the circuit will be transpiled to only basis gates, resulting in an even larger circuit. Combined, this makes it impossible to get a usable output when running this algorithm on a real quantum computer.
\\

We attempted to run these implementations on the real IBM Melbourne quantum computer. However, there is a limit for the circuit's execution time, and our transpiled circuit became too deep. This resulted in an exceeded time limit, and we could therefore not run the algorithm on the quantum device. 
\\

We also made some attempts to make the algorithm implementation hardware aware. The goal was to decrease the size of the transpiled circuit. We mainly focused on the hardware-aware implementation on the complete bipartite graph implementation since this graph had the smallest amount of qubits in the circuit and contained the least number of gates of the four implementations. The controlled QWALK gate was the main reason why the transpiled circuit was too deep; it was the sequence that used most multi-qubit gates. Most of these were a result of making the QWALK gate controlled. \\

As mentioned previously, two qubits can only interact with each other if they are directly connected. Since the Melbourne architecture allows a qubit to be directly connected to at most three other qubits, we cannot place all qubit that interacts via multi-qubit gates alongside each other. Therefore, we set the qubits that had the most interactions close, which reduced the number of swap gates in the transpiled circuit. In total, we succeeded in decreasing the QWALK gate by $12\%$.  However, this was not enough to make it executable on IBM's Melbourne computer. The main issue with the transpiled circuit was not the use of swap gates to position the qubits in the correct position; it was to transpile the circuit into basis gates. Too many basis gates were needed to represent the original circuit, and this is something that we cannot solve by adapting the circuit to the qubit architecture.

\section{Recommendation} \label{recommendation}
Since quantum computing is very far from ideal and suffers from much noise, one of the main improvements is to improve the hardware. As we have seen when running the implementation on a noisy simulator, the results would have been useless even if we managed to implement the algorithm on an actual quantum device.  For quantum computers to be usable in practice, the noise must decrease drastically.  One other crucial thing is to increase the number of qubits in the quantum computer to allow for even more extensive computations. There are also future improvements to the implementation of the quantum walk search algorithm, such as continuing to make it hardware-aware. We tried to make a subset of the implementation hardware-aware, but the next step is to do this with the complete implementation. As mentioned previously, the main issue was how the circuit was transpiled into the basis gates. One way of improving this might be to improve the algorithm that creates the transpiled circuit from the implemented circuits. One other way might be to manually implement the quantum walk search algorithm using these basis gates only. However, the most significant limitation of quantum computing today is the hardware, so continuing to work on these quantum devices and improving them is crucial for the future of quantum computing. \\

Another extension is to implement the quantum walk search algorithm with new types of graphs. This would require creating circuits for the new quantum walks. Also, the algorithm could be implemented with Szegedy's quantum walks, something that would also allow for irregular graphs.

\section{Conclusions} \label{conclusion}
In this paper, we implemented the quantum walk search algorithm. The implementation was done with four different regular graphs; a 4-dimensional hypercube graph, a 2-dimensional lattice graph, a complete bipartite graph, and a complete graph. We implemented the walks with a Grover coin, something that is equivalent to Szegedy's quantum walk. We executed the quantum walk search algorithm on both a quantum simulator without noise and a quantum simulator with noise. The result from running it on the simulator without noise matched the theoretical result, but when running it on the simulator with noise, this was not the case. This was caused by the error from using the gates, which increases when we increase the number of gates in the circuit. There were attempts to run the implementation on a real quantum computer; however, this was not possible since the transpiled circuits were too deep and required too long execution time. Although it is possible to implement the quantum search algorithm, which for an ideal quantum computer without noise will return the expected result, it is today impossible to run it on a real quantum computer and get a proper output. The error in the output will be too large, which makes the output unusable. The usefulness of implementing the quantum walk search algorithm will come first when quantum computers with less noise exist.

\clearpage
\appendix
\section{Quantum gates}
\subsection{Single qubit gates}
In this section, we present gates that operate on one qubit. 

\subsubsection{NOT gate} 
The NOT gate, or \textit{X} gate, inverts a qubit; $\ket{0}$ becomes $\ket{1}$ and $\ket{1}$ becomes $\ket{0}$. \\

Gate and matrix: \\

\begin{minipage}{0.3\textwidth}
\centering
\includegraphics[width=2cm, keepaspectratio=true]{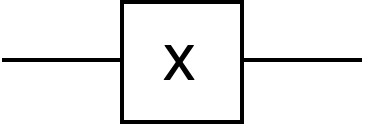}
\end{minipage}
\begin{minipage}{0.3\textwidth}
\centering
\includegraphics[width=2cm, keepaspectratio=true]{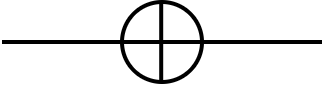}
\end{minipage}
\begin{minipage}{0.4\textwidth}
\centering
\begin{equation}
U_{\text{NOT}} =  \begin{bmatrix}
                    0 & 1 \\
                    1 & 0
\end{bmatrix}
\end{equation}
\null
\par\xdef\tpd{\the\prevdepth}
\end{minipage}

\subsubsection{Hadamard gate}
The Hadamard gate puts a qubit into equal superposition, i.e., after the gate has been applied, the probability that the qubit collapses to $\ket{0}$ or $\ket{1}$ are $50\%$ each. \\

Gate and matrix: \\

\begin{minipage}{0.5\textwidth}
\centering
\includegraphics[width=2cm, keepaspectratio=true]{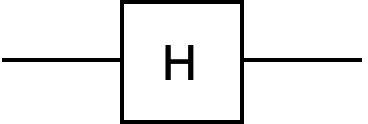}
\end{minipage}
\begin{minipage}{0.5\textwidth}
\centering
\begin{equation}
U_{\text{H}} = \frac{1}{\sqrt{2}} \begin{bmatrix}
                    1 & 1 \\
                    1 & -1
\end{bmatrix}
\end{equation}
\null
\par\xdef\tpd{\the\prevdepth}
\end{minipage}

\subsubsection{Z gate}
A Z gate shifts the phase of a qubit by $\pi$. It does not change state $\ket{0}$ but maps $\ket{1}$ to -$\ket{1}$. 

Gate and matrix: \\

\begin{minipage}{0.5\textwidth}
\centering
\includegraphics[width=2cm, keepaspectratio=true]{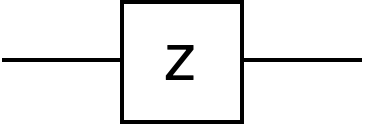}
\end{minipage}
\begin{minipage}{0.5\textwidth}
\centering
\begin{equation}
U_{\text{Z}} =  \begin{bmatrix}
                    1 & 0 \\
                    0 & -1
\end{bmatrix}
\end{equation}
\null
\par\xdef\tpd{\the\prevdepth}
\end{minipage}

\subsubsection{I gate}
An I gate, or \textit{identity gate}, leaves the qubit unaffected. That is, the qubit state is the same both before and after the gate. \\

Gate and matrix: \\

\begin{minipage}{0.5\textwidth}
\centering
\includegraphics[width=2cm, keepaspectratio=true]{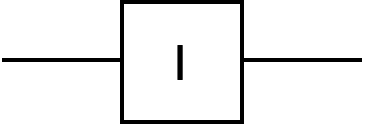}
\end{minipage}
\begin{minipage}{0.5\textwidth}
\centering
\begin{equation}
U_{\text{I}} = \ \begin{bmatrix}
                    1 & 0 \\
                    0 & 1
\end{bmatrix}
\end{equation}
\null
\par\xdef\tpd{\the\prevdepth}
\end{minipage}

\subsubsection{$R_Z$ gate}
The $R_Z$ gate, also referred to as $R_\phi$ gate, shifts the phase of qubits in state $\ket{1}$. The rotation angle $\phi$ is given as a parameter in the input and has to be a real number. \\

Gate and matrix: \\

\begin{minipage}{0.5\textwidth}
\centering
\includegraphics[width=2cm, keepaspectratio=true]{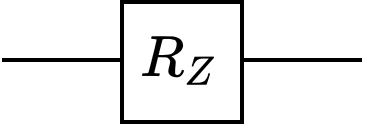}
\end{minipage}
\begin{minipage}{0.5\textwidth}
\centering
\begin{equation}
U_{\phi} = \ \begin{bmatrix}
                    1 & 0 \\
                    0 & e^{i \phi}
\end{bmatrix}
\end{equation}
\null
\par\xdef\tpd{\the\prevdepth}
\end{minipage}

\subsubsection{$\sqrt{X}$ gate}
$\sqrt{X}$, or $SX$ gate, is the square root of the NOT gate. \\

Gate and matrix: \\

\begin{minipage}{0.5\textwidth}
\centering
\includegraphics[width=2cm, keepaspectratio=true]{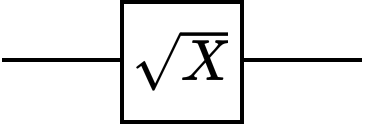}
\end{minipage}
\begin{minipage}{0.5\textwidth}
\centering
\begin{equation}
U_{\sqrt{X}} = \frac{1}{2} \begin{bmatrix}
                    1+i & 1-i \\
                    1-i & 1+i
\end{bmatrix}
\end{equation}
\null
\par\xdef\tpd{\the\prevdepth}
\end{minipage}

\subsection{Multi-qubit gates}
Below we introduce gates that act on two or more qubits. 

\subsubsection{CNOT gate}
A CNOT gate, or \textit{controlled-NOT gate}, operates on two qubits; one control qubit and one target. The gate is conditional; it applies a NOT gate on the target qubit if the control is $\ket{1}$. Otherwise, both qubits are left unchanged. \\

Gate and matrix: \\

\begin{minipage}{0.5\textwidth}
\centering
\includegraphics[width=3cm, keepaspectratio=true]{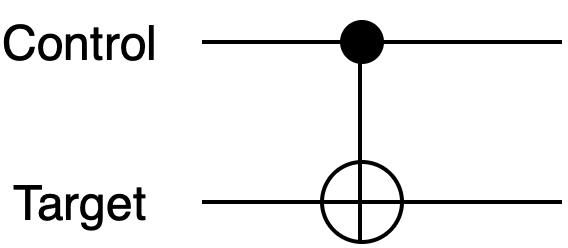}
\end{minipage}
\begin{minipage}{0.5\textwidth}
\centering
\begin{equation}
U_{\text{CNOT}} = \ \begin{bmatrix}
                    1 & 0 & 0 & 0 \\
                    0 & 1 & 0 & 0 \\
                    0 & 0 & 0 & 1 \\
                    0 & 0 & 1 & 0
\end{bmatrix}
\end{equation}
\null
\par\xdef\tpd{\the\prevdepth}
\end{minipage}

\subsubsection{$n$-bit Toffoli gate} \label{sec:n_toffoli}
We can generalize the idea of the CNOT gate to a gate with multiple control qubits. We call such a gate an \textit{$n$-bit Toffoli gate}, and it has $n-1$ control qubits and one target. A $2$-bit Toffoli gate is normally referred to as just \textit{Toffoli gate}, or \textit{CCNOT} gate. \\

Gate and matrix: \\ 

\begin{minipage}{0.5\textwidth}
\centering
\includegraphics[width=2cm, keepaspectratio=true]{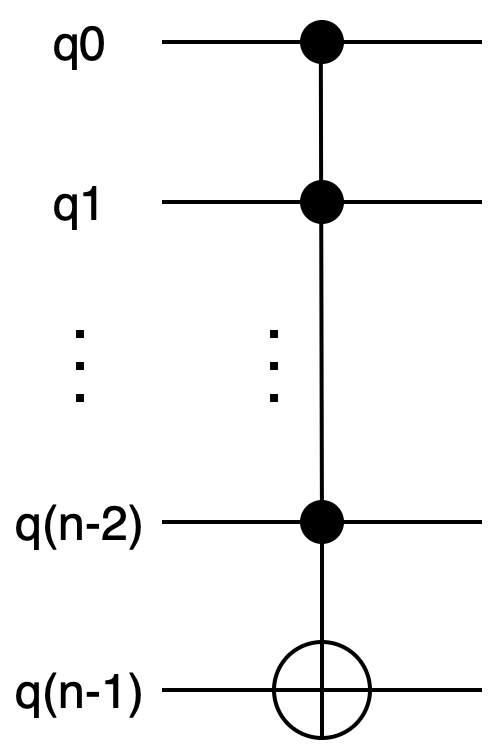}
\end{minipage}
\begin{minipage}{0.5\textwidth}
\centering
\begin{equation}
U_{\text{$n$-Toffoli}} = \begin{bmatrix}
                    1 & 0 & 0 & \dots & 0 & 0\\
                    0 & 1 & 0 & \dots & 0 & 0 \\
                    0 & 0 & \ddots  & \dots & 0 & 0 \\
                    \vdots  &  \vdots & \vdots & 1 & 0 & 0 \\
                    \vdots & \vdots & \vdots & \dots & 0 & 1 \\
                    0 & 0 & 0 & 0 & 1 & 0
\end{bmatrix}
\end{equation}
\null
\par\xdef\tpd{\the\prevdepth}
\end{minipage}

\subsubsection{SWAP gate}
A SWAP gate exchanges the values between two qubits so that the first qubit gets the state of the second and the other way around. \\

Gate and matrix: \\

\begin{minipage}{0.5\textwidth}
\centering
\includegraphics[width=1.5cm, keepaspectratio=true]{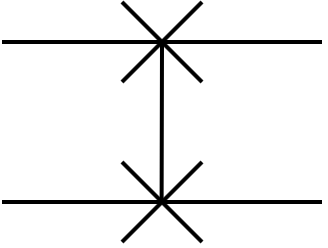}
\end{minipage}
\begin{minipage}{0.5\textwidth}
\centering
\begin{equation}
U_{\text{SWAP}} = \ \begin{bmatrix}
                    1 & 0 & 0 & 0\\
                    0 & 0 & 1 & 0 \\
                    0 & 1 & 0 & 0 \\
                    0 & 0 & 0 & 1
\end{bmatrix}
\end{equation}
\null
\par\xdef\tpd{\the\prevdepth}
\end{minipage}

\subsubsection{Controlled-Z gate}
Similarly to a controlled-NOT gate, we can construct a controlled-Z gate with either one or multiple control qubits and one target. Such a gate applies a Z-gate to the target if all control qubits are $\ket{1}$. The figures below show the matrix and gate for a controlled-Z gate with one control qubit, but we can extend it in the same way as we did with the CNOT gate in section \ref{sec:n_toffoli}. There are two ways to denote the control-Z gate, and both are shown below.  \\

Gate and matrix: \\

\begin{minipage}{0.3\textwidth}
\centering
\includegraphics[width=3cm, keepaspectratio=true]{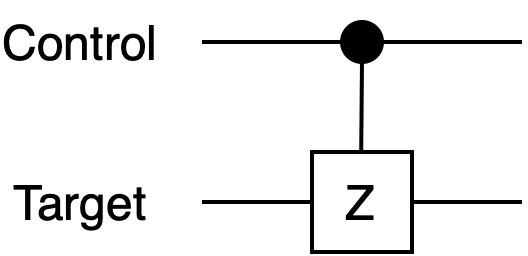}
\end{minipage}
\begin{minipage}{0.3\textwidth}
\centering
\includegraphics[width=3cm, keepaspectratio=true]{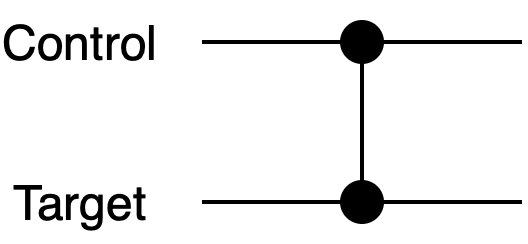}
\end{minipage}
\begin{minipage}{0.4\textwidth}
\centering
\begin{equation}
U_{\text{controlled-Z}} = \ \begin{bmatrix}
                    1 & 0 & 0 & 0\\
                    0 & 1 & 0 & 0 \\
                    0 & 0 & 1 & 0 \\
                    0 & 0 & 0 & -1
\end{bmatrix}
\end{equation}
\null
\par\xdef\tpd{\the\prevdepth}
\end{minipage}

\subsubsection{$\ket{0}$ as control qubit}
In all control gates we have presented above, we apply a gate to the target qubit if all control qubits are in state $\ket{1}$. However, we might want to operate on the target qubits if the control qubits are $\ket{0}$. We denote this by a white dot on the control qubits. The figure below shows this notation for a CNOT gate. 

\begin{figure}[h!]
\centering
\includegraphics[width=3cm]{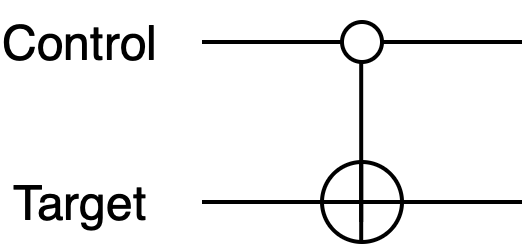}
\caption{Gate that applies a NOT gate on the target if the control is $\ket{0}$.}
\end{figure}

\clearpage
\printbibliography
\end{document}